\documentclass[aps,prd,twocolumn]{revtex4}
\usepackage[latin1]{inputenc}

\usepackage[english]{babel}
\usepackage{amsmath}
\usepackage{amsthm}
\usepackage{hhline}
\usepackage{amsfonts}
\usepackage{amssymb}
\usepackage{graphicx}

\theoremstyle{definition}

\usepackage{url}
\everymath{\displaystyle}
\usepackage{fancybox}

\usepackage{epsfig}
\usepackage{pgf}
\usepackage{verbatim}
\usepackage{listing}
\usepackage{xcolor}
\usepackage{fancyvrb}
\usepackage{multirow}
\usepackage{adjustbox}
\usepackage{tcolorbox}
\usepackage{enumitem}

\begin{document}

\title{Universality of the spherical collapse with respect to the matter type : the case of a barotropic fluid with linear equation of state}
\author{Fran\c{c}ois Staelens\textsuperscript{1}, J\'{e}r\'{e}my Rekier\textsuperscript{2}, Andr\'{e} F\"{u}zfa\textsuperscript{1}}
\email{francois.staelens@unamur.be}
\affiliation{\textsuperscript{1}Namur Institute for Complex Systems (naXys), University of Namur, Belgium\\
\textsuperscript{2}Royal Observatory of Belgium, Uccle, Belgium}

\date{\today}

\begin{abstract}
\noindent 
We study the spherical collapse of an over-density of a barotropic fluid with linear equation of state in a cosmological background. 
Fully relativistic simulations are performed by using the Baumgarte-Shibata-Shapiro-Nakamura formalism jointly with the Valencia formulation of the hydrodynamics. 
This permits us to test the universality of the critical collapse with respect to the matter type by considering the constant equation of state parameter $\omega$ as a control parameter. We exhibit, for a fixed radial profile of the energy-density contrast, the existence of a critical value $\omega^*$ for the equation of state parameter under which the fluctuation collapses to a black hole and above which it is diluting. It is shown numerically that the mass of the formed black hole, for subcritical solutions, obeys a scaling law $M\propto |\omega - \omega^*|^\gamma$ with a critical exponent $\gamma$ independent on the matter type, revealing the universality. This universal scaling law is shown to be verified in the empty Minkoswki and de Sitter space-times. For the full matter Einstein-de Sitter universe, the universality is not observed if conformally flat sub-horizon initial conditions are used. But when super-horizon initial conditions computed from the long-wavelength approximation are used, the universality appears to be true.
\end{abstract}


\maketitle
\section{Introduction}

The development of this $3+1$ formalism of General Relativity and the associated algorithmics during the $\text{XX}^\text{th}$ century combined with the "computer revolution" of the last decades permits the study of gravitation from a new point of view. The important works of Darmois (\cite{Darmois:1927aa}), Lichnerowicz (\cite{Lichnerowicz:1939aa}, \cite{Lichnerowicz:1944aa}, \cite{Lichnerowicz:1952aa}), Choquet-Bruhat (\cite{Foures:1956aa}), Dirac (\cite{Dirac:1958aa}, \cite{Dirac:1959aa}) and Arnowitt et al. (\cite{Arnowitt:1962aa}), managed to write Einstein equations as a constrained Cauchy problem, a suitable form for numerical integration, in the view of developing a quantum theory of gravitation. These laid the foundations of the Hamiltonian formulation of GR and introduced the well known Arnowitt-Deser-Misner (ADM) formalism. With the emergence of more and more powerful computers, numerical integration of such new formalisms became possible and the capability to simulate GR whatever the ingredients considered was a dream that scientists could then try to render realistic. This opened the era of Numerical Relativity, a new field of research whose aim is to build and use numerical methods to solve Einstein equations of GR on a computer. Remarkable works to mention are \cite{Nakamura:1987aa}, \cite{Nakamura:1994aa}, \cite{Shibata:1995aa}, \cite{Baumgarte:1998aa}, which developed the famous Baumgarte-Shibata-Shapiro-Nakamura (BSSN) formalism in the 1990's, the one used in this article.

Numerical Relativity obtained several successes and is now widely used in modern physics. It has been used, among others, to simulate spherical black holes formation (\cite{Shibata:1999aa}), stable solutions of neutron stars (\cite{Shibata:2002aa}) or binary black holes (\cite{Pretorius:2005aa}),\ldots The first detection of gravitational waves, the signal GW150914 from a binary black holes merger in 2016 by \cite{Abbott:2016aa} is an important evidence in favour of GR. This was made possible thanks to numerical relativity which permitted to verify post-Newtonian analytical developments and to go beyond it by simulating the black holes merger. This event has even enforced Numerical Relativity as an active, powerful and essential branch of physics. 

A second major result obtained thanks numerical relativity is the discovery of critical phenomena in gravitation. In 1992, Choptuik studied (see \cite{Choptuik:1993aa}) the spherical gravitational collapse of a massless scalar field thanks to numerical relativity. He found that, in a Minkowski background, the mass of a formed black hole $M$ follows a scaling power-law
\begin{equation}
M \propto (k-k_*)^\gamma,
\label{universal_powerlaw}
\end{equation}
where $k$ is a one dimensional quantity parametrising the initial data, $k_*$ is the threshold for black holes formation (which means that a black hole is formed when $k>k_*$ and not if $k<k_*$) and $\gamma$ is the critical exponent which does not depend on $k$. Critical phenomena following such a scaling law are said "universal". Moreover, the critical solution admits a continuous self-similarity (CSS). This critical phenomenon is similar to critical phase transitions found in statistical mechanic by identifying $M$ to an order parameter controlled by the function $|k-k_*|$ on the total phase space. 
 On this basis, numerous examples of universality in critical phenomena in gravitational collapse were discovered, some with a critical solution admitting a CSS and others with a discrete self-similarity (DSS). The interested reader can find more informations in the review by Gundlach and Martín-García \cite{Gundlach:2007aa}. Among others, it has been shown, still in a Minkowski background (\cite{Evans:1994aa}, \cite{Maison:1996aa}), that the universality was true in the case of the spherical collapse of a perfect fluid with barotropic equation of state $p = \omega e$, where $p$ is the pressure, $e$ is the energy density of the fluid and $\omega$ is a constant in the interval $[0,1]$. The associed critical solution is sometimes called the "Evans-Coleman" CSS solution, according to the authors of \cite{Evans:1994aa}. It was unclear if such CSS solutions exist for $\omega >0.89$ until \cite{Neilsen:2000aa} showed it was the case for all $\omega$ between $0$ and $1$. This discovery was of great importance because it means that, by fine tuning the initial conditions, it is possible to obtain a black hole with a mass as small as wished from a radiation fluid. The possible existence of tiny black holes would thus have an impact on the aboundance of primordial black holes formed during the radiation era. In 1999, \cite{Niemeyer:1999aa} performed simulations that showed that universality holds also in the cosmological case, when considering a non empty backgroung universe. However, \cite{Hawke:2002aa} showed in 2002, in a similar case, that some families of initial conditions admit a lower bound for the mass of a formed black hole : the scaling law \eqref{universal_powerlaw} did not work for values of $k$ very close to the critical solution but the mass seemed rather to stabilise towards $10^{-4}$ units of horizon mass. The authors explained that shocks, which are numerically challenging difficulties, are present when taking very small $|k-k_*|$ and this should be the reason why \cite{Niemeyer:1999aa} could not observe this phenomenon. 
 
Universality in gravitation collapse is thus a widely studied concept since the development of numerical relativity. Scaling laws similar to \eqref{universal_powerlaw} have been searched in many other situations such as charged black hole mass, angular momentum, coupled scalar field, higher dimensions,... (see \cite{Gundlach:2007aa}). In this article, we try to answer to the open question of the universality with respect to the matter type, in the case of a barotropic perfect fluid with linear equation of state $p=\omega e$. Indeed, at fixed initial data, varying the parameter $\omega$ will intuitively divide the solutions space into collapsing and non collapsing solutions, separated by a critical solution $\omega^*$ which should inevitably be the corresponding Evans-Coleman CSS solution. The idea is thus to see if $|\omega-\omega^*|$ can be considered as a control parameter, as well as $|k-k_*|$ was in previous cases, and if a similar scaling law is verified. 

To perform all this, we are following the works made in \cite{Rekier:2015aa} and \cite{Rekier:2016aa}, within the framework of numerical relativity. We use the BSSN formalism of GR, in spherical symmetry, conjointly with the Valencia formulation for the hydrodynamics \cite{Banyuls:1997aa}. Many numerical simulations in spherical symmetry use formalisms specially adapted to this kind of symmetry, such as the Misner-Sharp formalism (see \cite{Baumgarte:2010aa} or \cite{Shibata:2016aa} for a presentation of this formalism). A drawback of these formalisms is that they are written in comoving gauges. Because we intent to study in the future the spherical collapse of several fluids with relative velocities, comoving gauges were not suitable at the moment. The polar-areal gauge is often used too, such as in \cite{Noble:2016aa}, but is difficult to use in a cosmological context since it does not converge to the FLRW metric as $r\to \infty$. This is why we chose the BSSN formalism. When spherical coordinates are employed, it induces terms of the form $1/r^m$ which become problematic near the the origin of coordinates (i. e. when $r\to 0$). To overcome this difficulty, the authors of \cite{Cordero:2016aa} developped a partially implicit Runge-Kutta (PIRK) method for hyperbolic wave-like equations which solves the problems of instabilities without other regularization. This scheme has already been applied with success in the case of asymptotic flatness in \cite{Montero:2012aa}. For the case of an expanding background universe, it has also been done but only in the case of dust in \cite{Rekier:2015aa} and in the case of a scalar field in \cite{Rekier:2016aa}. We apply here to the case of pressured matter.

Our results exhibit the existence of a critical $\omega^*$ under which the solution collapses to a black hole and above which it dilutes into the background. Concerning the collapse case, we obtain a scaling law similar to \eqref{universal_powerlaw}, with $|\omega - \omega^*|$ as control parameter, in the Minkowski and the de Sitter cosmologies. In a full matter Einstein-de Sitter universe, we observe a breaking of the universality when using conformally flat initial conditions in the sub-horizon regime. But when using a super-horizon fluctuation with exclusively growing modes (the long-wavelength solution), the universal scaling law appears to be true.

This work is interesting from GR and mathematical points of view, but it has also important cosmological motivations. Indeed, the large scale structure formation mechanism is still not completely understood and our work could be a starting point for the study of spherically symmetric fluctuations evolution at several cosmological epochs. It could extend from primordial black holes formation in the radiation era to long term evolutions going through the equivalence radiation-dust epoch by considering two-fluids simulations.

We organize the paper as following. Section \ref{evolution eq} presents the BSSN formalism and all the equations that we use. The integration method, the gauge conditions and the choice of the initial data are described in section \ref{implementation}. The section \ref{initial_conditions} is devoted to the question of the initial conditions. The section \ref{results} contains all the numerical results, including the code validation. We give our conclusions and perspectives in the last section of the article.

\section{Evolution equations}
\label{evolution eq}

We give here a summary of the formalism we used to solve numerically Einstein and hydrodynamical equations. In all what follows, we work in natural units in which $G=c=1$. The time scale $t_\text{scale}$, in $\text{s}$, is fixed through  the comparison of the experimental value of the Hubble factor measured today $H_0^\text{exp} \sim 70\text{km} /\text{s}/\text{Mpc}$ with the adjustable parameter $H_0$ in arbitrary units, chosen for numerical reasons : 
\begin{equation*}
t_\text{scale} = \frac{H_0}{H_0^\text{exp}}.
\end{equation*}
The length scale $l_\text{scale}$, in $\text{m}$, and mass scale $m_\text{scale}$, in $\text{kg}$, are thus computed through
\begin{eqnarray*}
l_\text{scale} &=& ct_\text{scale} \\
m_\text{scale} &=& \frac{c^3}{G}t_\text{scale}.
\end{eqnarray*}

\subsection{BSSN formalism in spherical symmetry}

We follow what was made in \cite{Rekier:2015aa} and because of spherical symmetry, we write the metric line element as
\begin{equation}
ds^2 = -(\alpha^2-\beta^2)dt^2 + 2\beta dr dt + \psi^4a^2(t)\left(\hat{a}dr^2 + \hat{b}r^2d\Omega^2\right)
\label{metric}
\end{equation}
where $\alpha(t,r)$ is the lapse, $\beta(t,r)$ the radial component of the shift vector, $\hat{a}$ and $\hat{b}$ are the non-zero components of the diagonal conformal spatial $3$-metric. The conformal factor is thus $\psi\sqrt{a}$ and we have factored out the cosmological scale factor $a(t)$ which follows its own dynamics ruled by background dynamical equations. The BSSN formalism ensures that $\det\left(\hat{\gamma}_{\mu\nu} \right)= 1$, where $\hat{\gamma}_{\mu\nu}$ is the conformal $3$-metric, which translates to $\hat{a}\hat{b}^2 = 1$. We have split the extrinsic curvature into its trace $K$ and its conformally trace-free part $\hat{A}_{ij}$ :
\begin{equation}
K_{ij} = \frac{1}{3}\gamma_{ij}K + \psi^4a^2\hat{A}_{ij},
\label{curvature}
\end{equation}
where $\gamma_{ij}$ is the spatial $3$-metric. Spherical symmetry impose that $\hat{A}_{ij}$ has only two non-zero components $A_a := \hat{A}^r_r$ and $A_b := \hat{A}^\theta_\theta$. Since $\hat{A}_{ij}$ is traceless, we have that $A_a+2A_b = 0$.

The particularity of the BSSN scheme lies in the addition of the auxiliary $3$-vector $\hat{\Delta}^i$, which corresponds to the conformal connection :
\begin{equation}
\hat{\Delta}^i := \hat{\gamma}^{jk}\hat{\Gamma}^i_{jk} = -\partial_j\hat{\gamma}^{ij}.
\end{equation}
In spherical symmetry, the only non-zero component of this vector is (see \cite{Alcubierre:2011aa}):
\begin{equation}
\hat{\Delta}^r = \frac{1}{\hat{a}}\left[\frac{\partial_r\hat{a}}{2\hat{a}} - \frac{\partial_r\hat{b}}{\hat{b}} - \frac{2}{r}\left(1-\frac{\hat{a}}{\hat{b}}\right)\right].
\label{Delta}
\end{equation}
As in \cite{Rekier:2015aa}, we restrict to the zero shift case, $\beta = 0$, for simplicity.

The energy source terms measured by an Eulerian observer are expressed by the projections of the energy-momentum tensor $T^{\mu\nu}$ :
\begin{eqnarray}
E &=& n_{\mu} n_{\nu}T^{\mu\nu}, \\
j_i &=& -\gamma_{i\mu}n_\nu T^{\mu\nu},\\
S_{ij} &=&  \gamma_{i\mu}\gamma_{j\nu}T^{\mu\nu},
\end{eqnarray}
where $n_\mu = \left(-\alpha,0,0,0\right)$ is the four-vector field orthogonal to the spatial hypersurfaces. The tensor $T^{\mu\nu}$ of a perfect fluid can be written in function of the rest-mass density (or the particle number density) $\rho$, the specific enthalpy $h$, the pressure $p$ and the fluid $4$-velocity $u^\mu$ :
\begin{equation}
T^{\mu\nu} = \rho h u^\mu u^\nu + p g^{\mu\nu}.
\label{Tmunu}
\end{equation}
Spherical symmetry imposes that the only independent quantities are $E$, $j^r$, $S_a := S^r_r$ and $S_b := S^\theta_\theta$. 

Following \cite{Alcubierre:2011aa} and \cite{Rekier:2015aa}, the evolution equations for all the dynamical variables are :
\begin{eqnarray}
\partial_t \hat{a} &=& -2\alpha\hat{a}A_a, \label{BSSN1} \\
\partial_t\hat{b} &=& -2\alpha\hat{b}A_b, \label{BSSN2}\\
\partial_t\psi &=& -\frac{1}{6}\alpha\psi K - \frac{1}{2}\frac{\dot{a}}{a}\psi, \label{BSSN3}\\
\partial_t K &=& -\nabla^2\alpha + \alpha\left(A_a^2 + 2A_b^2+\frac{1}{3}K^2\right) \nonumber\\
&&+ 4\pi \alpha \left(E + S_a + 2S_b\right),\\
\partial_t A_a &=& -\left(\nabla^r\nabla_r\alpha - \frac{1}{3}\nabla^2\alpha\right) + \alpha\left(^{(3)}R^r_r - \frac{1}{3}^{(3)}R\right)\nonumber\\
&& + \alpha K A_a - \frac{16\pi}{3}\alpha\left(S_a-S_b\right),\\
\partial_t\hat{\Delta}^r &=& -\frac{2}{\hat{a}}\left(A_a\partial_r \alpha + \alpha\partial_r A_a\right) \nonumber\\
&&+ 2\alpha\left( A_a\hat{\Delta}^r - \frac{2}{r\hat{b}}\left(A_a -A_b\right)\right)\nonumber \\
 && + \frac{\xi\alpha}{\hat{a}}\Big[\partial_r A_a - \frac{2}{3}\partial_r K + 6A_a\frac{\partial_r\psi}{\psi}\nonumber \\
 &&+ \left(A_a-A_b\right)\left(\frac{2}{r}+\frac{\partial_r\hat{b}}{\hat{b}}\right) - 8\pi j_r\Big],\label{BSSN6}
\end{eqnarray}
together with the evolution of the scale factor $a(t)$ throug Friedmann equation. We ensure strong hyperbolicity of the equations by setting $\xi = 2$ (see \cite{Alcubierre:2008aa} ).

In this formalism, the Hamiltonian and momentum constraint equations read
\begin{eqnarray}
\mathcal{H} &\equiv & ^{(3)}R - \left(A_a^2+2A_b^2\right) + \frac{2}{3}K^2- 16\pi E = 0,\label{Hamiltonian}\\
\mathcal{M}^r &\equiv & \partial_r A_a - \frac{2}{3}\partial_r K + 6 A_a \frac{\partial_r\psi}{\psi} \nonumber\\
&&+ \left(A_a-A_b\right)\left(\frac{2}{r}+\frac{\partial_r\hat{b}}{\hat{b}}\right) - 8\pi j_r = 0.\label{Momentum}
\end{eqnarray}
Those two equations are used to monitor the reliability and the stability of the method, the so-called \textit{free evolution scheme}.

The behaviours of the variables near the origin satisfy the following parity conditions to ensure regularity of the terms in $\frac{1}{r}$ (see \cite{Alcubierre:2011aa}) :

\begin{eqnarray}
\alpha &\sim & \alpha_0 + O(r^2),\\
\hat{a} &\sim & \hat{a}_0 + O(r^2),\\
\hat{b} &\sim & \hat{b}_0 + O(r^2),\\
A_a &\sim & A_a^0 + O(r^2),\\
A_b &\sim & A_b^0 + O(r^2),\\
\hat{\Delta}^r &\sim & O(r^2),
\end{eqnarray}

where $\alpha_0$, $\hat{a}_0$, $\hat{b}_0$, $A_a^0$ and $A_b^0$ are functions of time exclusively. Moreover, terms in $(A_a-A_b)/r$ and $(1+\hat{a}/\hat{b})/r$ must also disappear near the origin, which means that the two following conditions, the flatness regularity condition, must occur :

\begin{eqnarray}
A_a - A_b &\sim & O(r^2) \Leftrightarrow A_a^0 = A_b^0,\\
\hat{a}-\hat{b} &\sim & O(r^2) \Leftrightarrow \hat{a}_0 = \hat{b}_0.
\end{eqnarray}

Numerically, it is challenging to build a code that verifies simultaneously the parity and flatness regularity conditions. Several articles discuss this issue (see \cite{Arbona:1999aa}, \cite{Alcubierre:2005aa} and \cite{Ruiz:2008aa}). On our side, we do not need to implement such regularization conditions thanks to the kind of numerical code we use (see section \ref{implementation}).

\subsection{Valencia formulation for relativistic hydrodynamics with a reference metric}

The evolution of the source terms can be written in a conservative form by using the Valencia formulation (see \cite{Banyuls:1997aa}) which ensures stability. To do this, we follow \cite{Rezzolla:2013aa} (and its notations) and we define the vector $\mathbf{U} = \sqrt{\gamma}\left(D,S_r,\tau\right)$ containing the conserved variables :
\begin{eqnarray}
D &=& \rho W, \\
S_r &=& \rho h W^2 v_r, \\
\tau &=& \rho h W^2 - p - D.
\end{eqnarray}
where $v_r$ is the physical $3$-velocity of the fluid for an Eulerian observer and $W$ is the Lorentz factor :
\begin{eqnarray}
v^r &=& \frac{u^r}{\alpha u^t},\\
W &=& \alpha u^t = \frac{1}{\sqrt{1-v_rv^r}},
\end{eqnarray}
with $v^r$ in units of $c$.

We point out that it is generally not possible to recover the primitive variables ($v_r$, $h$, $p$, $\rho$, \ldots) from the conserved ones ($D$, $S_r$, $\tau$) in an analytical way, except in few particular cases. A root-finding procedure must be used (see \cite{Rezzolla:2013aa}).\\
The hydrodynamical equations $\nabla_\mu T^{\mu\nu} = 0$, jointly with the baryon number conservation, thus read
\begin{equation}
\partial_t \mathbf{U} + \partial_r \mathbf{F}^r = \mathbf{S},
\label{eq hydro vectorielle}
\end{equation}
where the fluxes $\mathbf{F}^r$ are 
\begin{equation}
\mathbf{F}^r = \sqrt{-g}\begin{pmatrix}
D v^r \\
S_rv^r + p\\
\tau v^r + pv^r
\end{pmatrix},
\end{equation}
and the sources are\footnote{We point out here an apparent little typo in \cite{Montero:2012aa} where the terms with indices $\theta$ are missing.}
\begin{eqnarray}
&&\mathbf{S} = \sqrt{-g}\begin{pmatrix}
0\\
\frac{1}{2}T^{\mu\nu}\partial_r g_{\mu\nu}\\
-T^{\mu\nu}\nabla_\mu n_\nu
\end{pmatrix}\nonumber\\
&&= \sqrt{-g}\begin{pmatrix}
0\\
-\alpha T^{00}\partial_r \alpha + \frac{1}{2}T^{rr}\partial_r \gamma_{rr} + T^{\theta\theta}\partial_r \gamma_{\theta\theta}\\
-T^{0r}\partial_r \alpha + T^{rr}K_{rr} + 2T^{\theta\theta} K_{\theta\theta}
\end{pmatrix}.
\end{eqnarray}
We recall that those expressions are exact only in the case of spherical symmetry and vanishing shift ($\beta = 0$). General equations can be found in \cite{Banyuls:1997aa} and \cite{Montero:2012aa}.

Using these equations in this form will be problematic in the case of a non-constant background metric. Indeed, the asymptotic value of the vector $\mathbf{U}$ is not well defined in spherical coordinates because the term $\sqrt{\gamma}$ diverges when $r\to +\infty$. To overcome this difficulty, we use the reference metric approach presented in \cite{Montero:2014aa}. It consists in taking as new variables $\tilde{\mathbf{U}} = \sqrt{\frac{\gamma}{\tilde{\gamma}}} \left( D,S_r,\tau\right)$, where $\tilde{\gamma}_{ij}$ is a reference metric whose determinant $\tilde{\gamma}$ is constant in time. In this case, the new fluxes are 
\begin{equation}
\tilde{\mathbf{F}}^r = \alpha\sqrt{\frac{\gamma}{\tilde{\gamma}}}\begin{pmatrix}
D v^r \\
S_rv^r + p\\
\tau v^r + pv^r
\end{pmatrix},
\end{equation}
and the source terms are
\begin{eqnarray}
\tilde{\mathbf{S}} &&= \alpha\sqrt{\frac{\gamma}{\tilde{\gamma}}}\begin{pmatrix}
0\\
-\alpha T^{00}\partial_r \alpha + \frac{1}{2}T^{rr}\partial_r \gamma_{rr} + T^{\theta\theta}\partial_r \gamma_{\theta\theta}\\
-T^{0r}\partial_r \alpha + T^{rr}K_{rr} + 2T^{\theta\theta} K_{\theta\theta}
\end{pmatrix}\nonumber\\
\nonumber\\
&& + \begin{pmatrix}
- \tilde{\mathbf{F}}^r_D \tilde{\Gamma}^k_{rk} \\
\tilde{\mathbf{F}}^r_{S_r} \left( \tilde{\Gamma}^r_{rr}-\tilde{\Gamma}^k_{rk}\right) + \alpha\sqrt{\frac{\gamma}{\tilde{\gamma}}} p \left( \tilde{\Gamma}^\theta_{\theta r} +\tilde{\Gamma}^\phi_{\phi r}  \right) \\
- \tilde{\mathbf{F}}^r_{\tau} \tilde{\Gamma}^k_{kr}
\end{pmatrix}.
\end{eqnarray}
The choice for $\tilde{\gamma}_{ij}$ is the flat metric in spherical polar coordinates : $\tilde{\gamma}_{ij} = \text{diag} (1, r^2, r^2\sin^2\theta )$.
Our final source terms are thus, after evaluating the connection symbols,
\begin{equation}
\tilde{\mathbf{S}} = \alpha\sqrt{\frac{\gamma}{\tilde{\gamma}}}\begin{pmatrix}
0\\
-\alpha T^{00}\partial_r \alpha + \frac{1}{2}T^{rr}\partial_r \gamma_{rr} + T^{\theta\theta}\partial_r \gamma_{\theta\theta}\\
-T^{0r}\partial_r \alpha + T^{rr}K_{rr} + 2T^{\theta\theta} K_{\theta\theta}
\end{pmatrix}  - \frac{2}{r}\tilde{\mathbf{F}}^r.
\end{equation}
Note that this expression can be derived in a direct way by simply developing the term $$\partial_r\mathbf{F}^r =  \partial_r\left(r^2\sin\theta \tilde{\mathbf{F}}^r\right) =  2r\sin\theta \tilde{\mathbf{F}}^r + r^2\sin\theta \partial_r \tilde{\mathbf{F}}^r$$ and inserting it in  \eqref{eq hydro vectorielle}. \\
With that choice, our variables $\tilde{\mathbf{U}}$ are well defined at spatial infinity because 
\begin{equation*}
\alpha \sqrt{\frac{\gamma}{\tilde{\gamma}}} = \alpha \psi^6 a^3\to \overline{\alpha} a^3\text{\ \ \            as    \ \ \      } r \to \infty,
\end{equation*}
where $\overline{\alpha}$ is the asymptotic lapse.

The last question that remains to be treated is to know what happens when we consider several matter fields. The easiest way to proceed is to build a stress tensor $T^{\mu\nu}_{(k)}$ for each matter field $k$ and to sum them all to find the total stress tensor :
\begin{equation*}
T^{\mu\nu} = \sum_k T^{\mu\nu}_{(k)}.
\end{equation*}

The corresponding source terms can be summed in the same way :
\begin{eqnarray*}
E &=& \sum_k E_{(k)}, \\
j_r &=& \sum_k j_r^{(k)}, \\
S_a &=& \sum_k S^{(k)}_a, \\
S_b &=& \sum_k S^{(k)}_b.
\end{eqnarray*}

Concerning the hydrodynamics equations \eqref{eq hydro vectorielle}, it depends on the adopted coupling between the fluids. In this work, we assume that there is no coupling between the fluids : each of them is conserved, independently from the others. We thus have $\nabla_\mu T^{\mu\nu}_{(k)} = 0$ for all $k$ and one set of hydrodynamical equations for each matter field.

\subsection{Equation of state}

To close the system, we need an equation of state $f(p,\rho,\epsilon) = 0$, where $\epsilon = h-1-\frac{p}{\rho}$ is the specific internal energy, which will describe what kind of fluid we are using. If we want to simulate an ideal gas, the equation of state will be of the form
\begin{equation}
p = \rho \epsilon \left(\mathbf{\gamma} - 1\right),
\label{eos_ideal}
\end{equation}
where $\mathbf{\gamma}$ is the adiabatic index. For a polytropic fluid, the equation will be
\begin{equation}
p = \mathbf{K}\rho^\textbf{$\Gamma$},
\label{eos_polytropic}
\end{equation}
where $\mathbf{K}$ is the polytropic constant and $\textbf{$\Gamma$}$ is the polytropic exponent. Those two cases are widely used in numerical relativity simulations. However, in cosmology we often work with a linear barotropic equation of state :
\begin{equation}
p = \omega e,
\label{eos w}
\end{equation}
where $e = \rho(1+\epsilon) = \rho h - p$ is the energy density. This equation is the limit $\epsilon \gg 1$, with $\omega = \gamma -1$, for fluids which do not have a rest-mass density (such as radiation). The variable $\rho$ is therefore not used in this case. This equation of state has the advantage to give a simple (and analytical) formula to recover the primitive variables from the conserved ones (see Appendix A). The value $\omega = 0$ represents a pressure-less matter (dust) while the value $\omega = \frac{1}{3}$ states for a radiation fluid (relativistic particles). Finally, $\omega < -\frac{1}{3}$ is the condition for the growth of the universe to be accelerated and the simplest way to achieve it is to consider a cosmological constant in the Einstein equations, corresponding to a constant value $\omega = -1$. However, in this work we consider mostly universes not accelerated, only filled with matter fields with $\omega \in \left[0,1\right]$.

We here point out the fact that, when dealing with homogeneous cosmological spacetimes, the energy density is usually denoted by the letter $\rho$ in the literature. In our case we prefer to use the letter $e$ and use the letter $\rho$ to denote the rest-mass density.

Our code permits to have two different kinds of matter with two different equations of state. The two fluids are considered as non coupled and thus are separately conserved. For example, we can run a simulation with dust and radiation by choosing $p_1 = 0$ and $p_2 = e_2/3$.

\section{Implementation}
\label{implementation}

To solve the hydrodynamical and BSSN equations, we use the same method as in \cite{Rekier:2015aa} (and first developped in \cite{Montero:2012aa}). The radial dimension is discretised by a uniformally cell-centered grid. A fourth-order finite difference scheme is used to compute radial derivatives and we use fourth-order Kreiss-Oliger dissipation. A few virtual points of negative radius are added to the grid to improve stability for the radial derivatives close to the origin by using parity conditions on the fields.

We use the PIRK methods to solve the evolution equations. To achieve it, we split the set of equations in two parts : 
\begin{equation}
\begin{cases}
\partial_t u = \mathcal{L}_1 (u,v), \\
\partial_t v = \mathcal{L}_2 (u) + \mathcal{L}_3 (u,v).
\end{cases}
\label{PIRK operators}
\end{equation}
The variables $u$ are first explicitly evolved and the result is used to evolve $v$ partially implicitly through the operator $\mathcal{L}_2$. Since it is a second order PIRK method, the evolution requires two steps which are described in details in \cite{Montero:2012aa}. In particular, if we denote by $L_1$, $L_2$ and $L_3$ the corresponding discrete operators of $\mathcal{L}_1$, $\mathcal{L}_2$ and $\mathcal{L}_3$ (the exact expressions for the splitting operators are given in Appendix B.), the operators $L_1$ and $L_3$ are used in an explicit way, while $L_2$ contains the unstable terms and is treated in a partially implicit way. This method has already been used in the frame of BSSN formalism under asymptotically flatness assumption (see \cite{Baumgarte:2013aa}). It has also been applied for a dynamical cosmological background in \cite{Rekier:2015aa} and \cite{Rekier:2016aa}, but it was restricted to the case of dust matter (pressureless matter) and scalar field, and so it did not really include hydrodynamics. The variables which are first explicitly evolved (those contained in the vector $u$) are the hydrodynamical conserved variables, the cosmological scale factor $a$, the lapse $\alpha$, the elements of the conformal $3$-metric $\hat{a}$ and $\hat{b}$ as well as $\psi$. Their updated values are subsequently used to evolve $K$, $A_a$ and $\hat{\Delta}^r$. 

Concerning the hydrodynamical equations, we first use a monotonised central-difference (MC) slope limiter (see \cite{Vanleer:1977aa}) to approximate the left and right states of the primitive variables at each cell. Secondly we solve the equations with a HLLE incomplete Riemann solver (from Harten, Lax, van Leer \cite{Harten:1983aa} and Einfeldt \cite{Einfeldt:1988aa}). Finally, we use a root-finding procedure (Newton-Raphson) to recover the primitive variables from the conserved ones if the equation of state is different than \eqref{eos w}.

\subsection{Gauge conditions}
\subsubsection{Background evolution}
As we said, we consider models in which space-time is not asymptotically constant but rather looks like a homogeneous FLRW Universe, without curvature, at large radii. Note that in all what follows, an overline is used to indicate the background value of the quantity.  The line-element is given by the FLRW metric, with possibly a dynamical lapse :

\begin{equation}
ds^2 = -\overline{\alpha}^2(t) dt^2 + a^2(t) \left( dr^2 + r^2d\Omega^2\right),
\end{equation}
where we consider that there is no curvature. The evolution of this metric is ruled by the well known Friedmann equations
\begin{eqnarray}
\frac{1}{\overline{\alpha}^2} \left(\frac{\dot{a}}{a}\right)^2 &=& \frac{8\pi}{3} \overline{e},\label{H_square_relation}\\
\frac{1}{\overline{\alpha}^2} \frac{\ddot{a}}{a} - \frac{\dot{a}}{a} \frac{\dot{\overline{\alpha}}}{\overline{\alpha}^3} &=& -\frac{4\pi}{6}\left( \overline{e} + 3\overline{p}\right),
\end{eqnarray}
where $\overline{e}$ and $\overline{p}$ are the homogeneous background energy density and pressure. These two quantities are composed with the contribution of different kinds of energy (matter (in general several species) and possibly a cosmological constant $\Lambda$ in our case) : 
\begin{eqnarray}
\begin{cases}
\overline{e} = \sum_k {\overline{e}_{k}}  + e_\Lambda, \\
\overline{p} =  \sum_k {\overline{p}_{k}} + p_\Lambda.
\end{cases}
\end{eqnarray}
For the background, the hydrodynamical equations are simplified and the only evolution equations that are needed are the following ones for the rest-mass density and the energy density :
\begin{eqnarray*}
\partial_t {\overline{\rho}_{k}} &=& -3\frac{\dot{a}}{a}{\overline{\rho}_{k}} = \overline{\alpha} \overline{K} {\overline{\rho}_{k}},\\
\partial_t {\overline{e}_k} &=& -3\frac{\dot{a}}{a}(\overline{e}_{k}+\overline{p}_k) = \overline{\alpha} \overline{K} (\overline{e}_{k}+\overline{p}_k),
\end{eqnarray*}
where $\overline{K} = -\frac{3}{\overline{\alpha}}\frac{\dot{a}}{a}$ is the trace of the homogeneous extrinsic curvature. Indeed, the other hydrodynamical variables can be recovered by using only the equation of state because the velocity ${{\overline{v}_k}^r}$ is null and the Lorentz factor is thus equal to $1$.
\subsubsection{Boundary conditions}

The spatial domain is of the form $r\in [0, r_{\text{span}} ]$, where $0$ corresponds to the origin and $r_{\text{span}}$ to the outer boundary. We use a cell-centered discretization to avoid calculations at the exact origin in case of singularities. At the origin, we impose, following spherical symmetry, the inhomogeneous variables to have the correct parity for a regular solution thanks to a few virtual points of negative radius we added to the grid. At the outer boundary, we use a Sommerfeld (radiative) boundary conditions (see \cite{Alcubierre:2003aa}) : we impose the variables to behave like outward travelling waves when $r$ is near $r_{\text{span}}$. This means that, at a few outermost points of the computational grid, any field $f(t,r)$ must verify 
\begin{equation}
\partial_t f = \partial_t \overline{f} - v\partial_r f - \frac{v}{r}\left( f-\overline{f}\right),\label{sommerfeld}
\end{equation}
where $v$ is the characteristic velocity of the field. Note that $v$ is computed by examining the dynamical equation of each field and is the speed of light for most of it. Only the lapse $\alpha$ and the variable $\hat{\Delta}^r$ admit a characteristic velocity different from it. For the lapse, it depends on the slicing that is used while it is $\sqrt{2}$ for $\hat{\Delta}^r$. Such a condition prevents any signal to be reflected by the outer boundary. The asymptotic values of each variables are the homogeneous ones given by the background evolution :
\begin{eqnarray}
\alpha(t,r) &\to & \overline{\alpha}(t), \\
\hat{a}(t,r),\hat{b}(t,r),\psi(t,r) &\to & 1, \\
K(t,r) &\to & \overline{K}(t) = -\frac{3}{\overline{\alpha}}\frac{\dot{a}}{a}, \\
A_a(t,r), A_b(t,r), \hat{\Delta}^r(t,r) &\to & 0,\\
\rho_{k}(t,r) &\to & {\overline{\rho}_{k}} (t), \\
e_{k}(t,r) &\to & {\overline{e}_{k}}(t), \\
{v_k}^r(t,r) &\to & 0.
\end{eqnarray}

\subsubsection{Slicing conditions}
\label{slicing_conditions}

There are lots of different slicing conditions in the literature (see for example \cite{Gourgoulhon:2012aa}, \cite{Baumgarte:2010aa} or \cite{Shibata:2016aa}). We implemented the Bona-Masso slicing (\cite{Bona:1995aa}) for the local dynamics and the geodesic slicing (constant unity lapse) for the background :
\begin{eqnarray}
\partial_t \alpha &=& -\alpha^2 f(\alpha) \left(K - \overline{K}\right),\label{slicing} \\
\overline{\alpha} &=& 1.
\end{eqnarray}
We differ a bit from what was made in \cite{Rekier:2015aa} and \cite{Rekier:2016aa}. They considered a slicing condition that did not converge to the geodesic one at spatial infinity. Our slicing conditions are thus not exactly the same, we added the $-\overline{K}$ in our equation to ensure that $\alpha \to 1$ as $r\to \infty$. This allows identifying the time coordinate to the cosmological synchronous time for physical interpretations. Such a slicing condition gives a characteristic velocity for the lapse of $\alpha\sqrt{f(\alpha)\gamma^{rr}}$ (see \cite{Alcubierre:2008aa}), which is equal to $\frac{\sqrt{f(\alpha =1)}}{a}$ at spatial infinity. This quantity must then be taken in the Sommerfeld conditions of \eqref{sommerfeld}.

Choosing $f\leq \frac{1}{3}$  implies that the coordinate speed of light remains finite (see \cite{Torres:2014aa} ). Thus, this condition ensures to keep the stability of the scheme though not mandatory. 

The simplest choice $f = 0$ is the geodesic slicing $\alpha = 1$. Combined with a zero shift $\beta = 0$ gives what is called the \textit{synchronous gauge}. Although this is not the best choice in term of stability - it easily generates coordinate singularities (see \cite{Gourgoulhon:2012aa})- we chose this one in some of our simulations because of its simplicity. 

To perform the simulation of black holes formation, we use the $1+\log$ slicing defined by $f(\alpha) = \frac{2}{\alpha}$. This seemed to be to most efficient Bona-Masso slicing in this context.

\subsection{Used quantities}
\label{used_quantities}

 In general relativity, deriving a well-posed definition of mass is a difficult challenge since it should be gauge invariant and, if possible, time invariant. Moreover, several notions of mass exist in general relativity.  For asymptotically Minskowski spacetimes, there still exists the possibility to compute the total energy on a single slice $\Sigma_t$. The ADM mass (see \cite{Shibata:2016aa})  is based on this principle and gives a time independent quantity. Another mass notion that is often used in static spacetimes is the Komar mass (see \cite{Gourgoulhon:2012aa}). But since we are working in an asymptotically Friedmann universe, we cannot use those definitions with no change. Indeed, the total mass is not finite and we need to truncate the computation to keep only the mass inside the fluctuation. To achieve that, we follow \cite{Shibata:1999aa} and \cite{Harada:2015aa} and use the Kodama mass, which was first define in \cite{Kodama:1980aa}. The definition et expression of this mass is given in details in Appendix C.
 Now that we have defined our mass notion, we are ready to use it to define a compactness notion. Usually, the compactness of an object is a dimensionless quantity defined as its ratio mass to radius : $\frac{G}{c^2}\frac{M}{R}$. Similarly, we define the compaction function as the mass excess inside the sphere with areal radius $R$ :
\begin{eqnarray}
C(t,r) = \frac{2\left(M_K(t,r)-\overline{M_K}(t,\psi^2\sqrt{\hat{b}}r)\right)}{R},
\label{compact_definition}
\end{eqnarray}
where the factor $2$ is simply a question of convention to follow the definition of \cite{Musco:2019aa}. This quantity is useful to define the size of the fluctuation. The radius $r_m(t)$ is defined, in hand, as the coordinate $r$ where the compaction function reaches its maximal value. With this point, we can define the mass of the fluctuation as
\begin{equation}
M_m(t) := M_K(t,r_m(t))
\end{equation}
 and its compactness as 
 \begin{equation}
 C_m(t) := C(t,r_m(t)) = \max_{r>0}C(t,r).
 \end{equation}

The last quantity we need before using the code is the amplitude of the fluctuation. The standard way in cosmology is to take the central value of the energy density contrast. But in fact, there is no reason to consider only this particular value because this quantity is not necessarily representative of the full behaviour of the fluctuation, especially when pressure enters into consideration. Indeed, the energy-density contrast radial profile changes when pressure increases. This is why we use an average energy density contrast defined by
\begin{equation}
\overline{\delta}(t,r) = \frac{\int^R_0 4\pi \delta R'^2dR'}{\int^R_0 4\pi R'^2dR'}.
\end{equation}

The mean energy-density contrast of the fluctuation, i.e. its amplitude, is defined as this quantity evaluated at the radius of the fluctuation : 
\begin{equation}
\overline{\delta}_m(t) = \overline{\delta}(t,r_m(t)).
\end{equation}
The pertinence of this definition can be pointed out through the following relations obtained in the long-wavelength approximation :
\begin{eqnarray}
{\overline{\delta}_m}(t) &\simeq & 3 \delta(t,r_m(t))\label{relation_musco} \\
C_m(t) &\simeq & {\overline{\delta}_m}(t) \left( H(t)R(t,r_m(t))\right)^2,\label{relation_harada}
\end{eqnarray}
where the first approximation has been derived in \cite{Musco:2019aa} and the second in \cite{Harada:2015aa}. These relations show that the particular radius $r_m(t)$ contains an important part of the information characterizing the fluctuation whatever its shape.

To conclude, the quantities we will focus on are the following ones:
\begin{itemize}
\item[-] the central energy density contrast $\delta_0(t)$ ;
\item[-] the mean energy density contrast $\overline{\delta}_m(t)$ ;
\item[-] the compactness $C_m(t)$ ;
\item[-] the radius $r_m(t)$ and the corresponding areal radius $R_m(t)$ ;
\item[-] the mass $M_m(t)$ ; 
\item[-] the central value of the lapse $\alpha_0(t)$ when a non geodesic slicing is used.
\end{itemize}

\section{Initial conditions}
\label{initial_conditions}

There are two ways of setting the initial conditions : either we choose well determined values for almost all the variables and derive the last ones with the constraint equations \eqref{Hamiltonian} and \eqref{Momentum}, or we choose initial conditions following a good approximation of the solution modelling the situation at early times. The first method has the advantage to start with, theoretically, an exact solution of general relativity since the constraint equations are verified at initial time. But the initial data generated in this way can be viewed as quite artificial and could suffer from transient effects at early times before reaching a more realistic evolution. The second method has opposite properties. It generates a more realistic initial situation from the physical point of view but this is generally only an approximation of the exact solution and then needs some iterations for the constraints to reach a more acceptable level of accuracy. We start by exhibiting the first method in a simple but useful case.

\subsection{Conformally flat initial conditions}
\label{conformally_flat}

This idea here is to assume spatial homogeneity on each metric variables except the conformal factor and thus to equal them to their corresponding background values :

\begin{eqnarray}
\alpha (t=0,r) &=& 1,\\
\hat{a}(t=0,r) = \hat{b}(t=0) &=& 1,\\
K(t=0,r) &=& \overline{K}_i = -3H_i,\\
A_a(t=0,r) = A_b(t=0,r) &=& 0,\\
\hat{\Delta}^r(t=0,r) &=& 0,
\end{eqnarray}
where $\overline{K}_i$ is the initial background curvature and $H_i$ the initial Hubble factor. 

Concerning the matter source terms, the Momentum constraint \eqref{Momentum} reduces to $j^r(t=0,r) = 0$. This imposes a null spatial $3-$velocity :
\begin{equation*}
v^r_k(t=0,r) = 0.
\end{equation*}
Note that in the case of several fluids, the condition $j^r(t=0,r) = 0$ is weaker than $v^r_k(t=0,r) = 0$ for all $k$. But other possibilities are much more complicated and, we think, naturally improbable.

Since we will work with the cosmological equation of state $p=\omega e$, we recall that we do not have to use the rest-mass density quantity $\rho$. Only two last quantities thus remain to be specified : the energy density and the initial curvature, i.e. the conformal factor. Both are linked by the Hamiltonian constraint :
\begin{equation}
\label{bvp_equation}
-a^{-2}\psi^{-5}\left(\partial^2_r\psi + \frac{2}{r}\partial_r \psi\right) + \frac{3}{4} H^2_i = 2\pi E(t=0,r).
\end{equation}

So, specifying one determines the other. Intuitively, the physicist prefers to introduce a specific energy-density profile to model a situation. This is what was done in \cite{Rekier:2015aa}, \cite{Rekier:2016aa} and \cite{Shibata:1999aa}. The initial conformal factor $\psi(t=0,r)$ is then found by solving numerically \eqref{bvp_equation} as a boundary value problem with
\begin{eqnarray}
\partial_r \psi \to &  0,&\hspace{2cm}\text{for $r\to 0$;}\\
\psi \to & 1+\frac{C_\psi}{2r},&\hspace{2cm}\text{for $r\to \infty$},
\end{eqnarray}
where $C_\psi$ is a constant adjusted such that
\begin{eqnarray}
\partial_r \psi \to & -\frac{C_\psi}{2r^2},&\hspace{2cm}\text{for $r\to \infty$.}
\end{eqnarray}

This method is used successfully in section \ref{validation} and in all the following chapter. We specify the initial energy density in terms of the energy density contrast $\displaystyle \delta^i_k(r) = \frac{e_k(t=0,r)}{\overline{e}_k(t=0)}-1$. We use a smooth top-hat profile (a logistic function) to simulate the evolution of an over-dense region of the universe :
\begin{equation}
\delta^i_{k} (r) = \delta^i_{k} \frac{1-\tanh\left(\frac{r-{r^i_{k}}}{2\sigma_{k}} \right) }{1+\tanh\left(\frac{r^i_{k}}{2\sigma_{k}}\right)},
\label{initial_profile}
\end{equation}
where $\delta^i_{k}$, ${r^i_{k}}$ and $\sigma_{k}$ are positive parameters designing the shape of the profile.

\subsection{The long-wavelength approximation}
\label{sec:long_walength}

To generate more realistic initial conditions, we can use the long-wavelength approach (also called \textit{gradient expansion}), as described in \cite{Lyth:2005aa}, \cite{Harada:2015aa} and \cite{Musco:2019aa}. This approach resembles the cosmological perturbation theory but, instead of developing the equations in powers of the inhomogeneities as in the lattice, the long-wavelength scheme expands the solution in the spatial gradient of these perturbations. Concretely, we focus on some fixed time, multiply each spatial gradient $\partial_i$ by a fictitious parameter $\epsilon \ll 1$ and we expand the equations in a power series of $\epsilon$. We keep terms up to first-order in $\epsilon$ and then set $\epsilon = 1$. To compare, in the perturbative approach the parameter $\epsilon$ would multiply the perturbation instead of $\partial_i$. 

The parameter $\epsilon$ is conveniently identified as the ratio
\begin{equation}
\epsilon := \frac{R_H(t)}{L},
\end{equation}
where $\displaystyle R_H(t) = \frac{1}{H(t)}$ is the Hubble radius, the only geometrical scale in the homogeneous universe, and $L$ is the (comoving) length scale of the perturbation. This approach reproduces the results of linear perturbation theory but can also consider non linear perturbations of the curvature if the universe is sufficiently smooth for scales greater than $L$ (see \cite{Lyth:2005aa} and \cite{Musco:2019aa}).

The complete derivation of the solution in a general Bona-Masso gauge \eqref{slicing} is presented in the Appendix D. In terms of BSSN variables in spherical symmetry, the long-wavelength solution is given by
\begin{eqnarray}
\psi &=& \Psi \left(1 -\frac{1}{6(1+\omega )} \frac{v_*^2}{v_*^2-2v_*^1}F \left(\frac{1}{aH}\right)^2\right) \nonumber\\
&&+ O(\epsilon^4),\\
A_a &=& \frac{2}{3\omega +5}p_a H \left(\frac{1}{aH}\right)^2 + O(\epsilon^4),\\
A_b &=& \frac{2}{3\omega +5}p_b H \left(\frac{1}{aH}\right)^2 + O(\epsilon^4),\\
\hat{a} &=& 1 - \frac{4}{(3\omega +5)(3\omega+1)}p_a\left(\frac{1}{aH}\right)^2 + O(\epsilon^4),\\
\hat{b} &=& 1 - \frac{4}{(3\omega +5)(3\omega+1)}p_b\left(\frac{1}{aH}\right)^2 + O(\epsilon^4),\\
\delta &=& \frac{v_*^2}{v_*^2-2v_*^1}F \left(\frac{1}{aH}\right)^2 + O(\epsilon^4),\label{delta_lwa} \\
K &=& \overline{K}\left(1+ \frac{v_*^1}{v_*^2-2v_*^1}F \left(\frac{1}{aH}\right)^2\right) +O(\epsilon^4),\\
\alpha &=& 1+\frac{v_*^3}{v_*^2-2v_*^1}F \left(\frac{1}{aH}\right)^2 + O(\epsilon^4),\\
v_r &=& -\frac{2}{3\omega+5}\cdot\frac{\omega v_*^2+(1+\omega) v_*^3}{(1+\omega)(v_*^2-2v_*^1)}\partial_i F a \left(\frac{1}{aH}\right)^3 \nonumber\\
&&+ O(\epsilon^5).
\end{eqnarray}
where 
\begin{eqnarray}
\begin{pmatrix}
v_*^1\\
v_*^2\\
v_*^3
\end{pmatrix} = \begin{pmatrix}
(1+3\omega )^2\\
-3(1+\omega)(1+3\omega+3f(1))\\
3(1+3\omega)f(1)
\end{pmatrix}.
\end{eqnarray}
 and we used the intermediate variables $p_a$, $p_b$ and $F$ given by
\begin{eqnarray}
p_a &:=& \frac{1}{\Psi^4}\left[-\frac{4}{3\Psi}\left(\partial^2_r\Psi -\frac{1}{r}\partial_r\Psi\right) + \frac{4}{\Psi^2}\left(
\partial_r\Psi\right)^2\right]\hspace*{-1mm},\\
p_b &:=&  \frac{1}{\Psi^4}\left[\frac{2}{3\Psi}\left(\partial^2_r\Psi -\frac{1}{r}\partial_r\Psi\right) - \frac{2}{\Psi^2}\left(
\partial_r\Psi\right)^2\right],\\
F &:=& -\frac{4}{3} \frac{\overline{\Delta \Psi}}{\Psi^5}.
\end{eqnarray}

If we write
\begin{equation}
\Psi := e^{\frac{\zeta}{2}},
\end{equation}
we have the following expressions for $p_a$, $p_b$ and $F$ :
\begin{eqnarray}
p_a &=& -\frac{2}{3}e^{-2\zeta}\left( \zeta'' + \zeta'\left(\frac{1}{r}-\zeta'\right)\right),\\
p_b &=& \frac{1}{3}e^{-2\zeta}\left( \zeta'' + \zeta'\left(\frac{1}{r}-\zeta'\right)\right),\\
F &=& -\frac{2}{3}e^{-2\zeta}\left(\zeta'' +\zeta'\left(\frac{2}{r}+\frac{\zeta'}{2}\right)\right).
\end{eqnarray}
Moreover, if we choose the Gaussian profile for $\zeta$, i.e.
\begin{equation}
\zeta(r) = A\exp\left(-\frac{r^2}{2\Delta^2}\right),
\label{profile_zeta}
\end{equation}
these expressions become
\begin{eqnarray}
p_a &=& \frac{4\zeta}{3\Delta^2}e^{-2\zeta} \left(1-\frac{r^2}{2\Delta^2}\left(1-\zeta\right)\right),\\
p_b &=& -\frac{2\zeta}{3\Delta^2}e^{-2\zeta} \left(1-\frac{r^2}{2\Delta^2}\left(1-\zeta\right)\right),\\
F &=& \frac{2\zeta}{\Delta^2}e^{-2\zeta} \left(1-\frac{r^2}{3\Delta^2}\left(1+\frac{\zeta}{2}\right)\right).
\end{eqnarray}

As a confirmation of the development, the particular case of the geodesic slicing with $f(\alpha) = 0$ gives a solution that is identical to the one developed in \cite{Harada:2015aa}. This solution will give realistic initial conditions for the spherical collapse if the length-scale of the inhomogeneity is greater than the Hubble radius.

\section{Numerical results}
\label{results}

We start now the section which presents our all our numerical results. To avoid the reader to be lost, we give here a little explanation of the process presented in this section. We first validate the code in section \ref{validation}. Then, we present, in section \ref{section_behaviour}, the two typical behaviours that we observed in our simulations : the collapse and dilution scenarios. The next step is the choice of a gauge that permits to simulate black holes formations. This is done in section\ref{sec:up_and_down_simulation} where simulations of sub and super-critical solutions are performed. Finally, the universality of the phenomenon is investigated in section \ref{universality} with this specific gauge.

\subsection{Code validation}
\label{validation}

To achieve this validation, we performed a simulation with two species of matter which have linear equations of state $p_{1} = 0.1 e_{1}$ and $p_{2} = 0$. We chose the geodesic slicing and the initial data are fixed by the conformally flat method discussed in section \ref{conformally_flat}. Note that we do not validate our code with the long-wavelength approximation as initial conditions since we need initial conditions perfectly correct (not an approximation) from the general relativity point of view. The initial profiles for the energy density contrasts $\delta_{k} = \frac{e_{k}}{{\overline{e}_{k}}} -1$ are of the form of the smooth top-hat functions given by eq. \eqref{initial_profile}. We fix the positive parameters of these profiles to $\delta^i_{1} = \delta^i_{2} = 1$, ${r^i_{1}} = {r^i_{2}} = 10$ and $\sigma_{1} = \sigma_{2} = 1$, while our spatial domain is the interval $[ 0,500]$ (all in code units). We put as much quantity of matter $1$ as of matter $2$, which reads $\Omega^i_{1} = \Omega^i_{2} = 0.5$.

The last quantities that remain to be fixed are the initial scale factor $a_i$, the initial Hubble factor $H_i$ and the Hubble factor measured today $H_0$ which will determine the time scale, the mass scale and the length scale. For our tests, we chose $a_i = 1$, $H_i = 0.03$ and $H_0 = 0.001$. The Courant-Friedrichs-Lewy factor (CFL) is set to $0.25$, indicating that the discretization step in time $\Delta t$ is linked to the spatial discretization $\Delta r$ through the relation $\Delta t = 0.25\Delta r$. We tested the three resolutions $\Delta r = 0.1$, $\Delta r = 0.05$, and $\Delta r = 0.025$.

\begin{figure}[!]
\centering
	\includegraphics[height = 7cm]{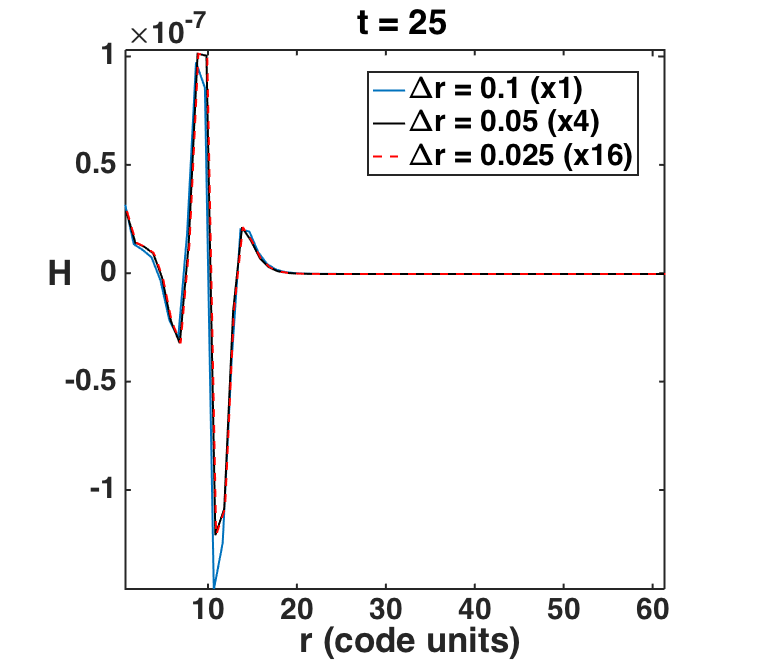} 
    \caption{\footnotesize{Hamiltonian constraint at $t = 25$ for simulation of the evolution of a smooth inhomogeneity in the density profile, with three resolutions : $\Delta r = 0.1$, $\Delta r = 0.05$, and $\Delta r = 0.025$. Curve for $\Delta r = 0.05$ has been multiplied by $4$ and curve for $\Delta r = 0.025$ has been multiplied by $16$ to exhibit the second order of convergence of the method.}}
    \label{hamilt_t25}
\end{figure}

We now present the results of these simulations. The Hamiltonian constraint at $t = 25$ is shown on Fig. \ref{hamilt_t25}. The similarity in the shapes of the curves and the fact that it is rescaling with the resolution in the right order (curve for $\Delta r = 0.05$ has been multiplied by $4$ and curve for $\Delta r = 0.025$ has been multiplied by $16$) show the stability of the method and at least a second-order convergence of the scheme. Of course, the error is maximal at the centre of coordinates and at the boundary between the inner and outer parts of the over-density. Terms in inverse power laws of the radius are responsible for the larger error at the center. For the overdensity boundary , it is the location where the gradients are maximal and it justifies these peaks in the error.

\begin{figure}[!]
\centering
	\includegraphics[height = 7cm]{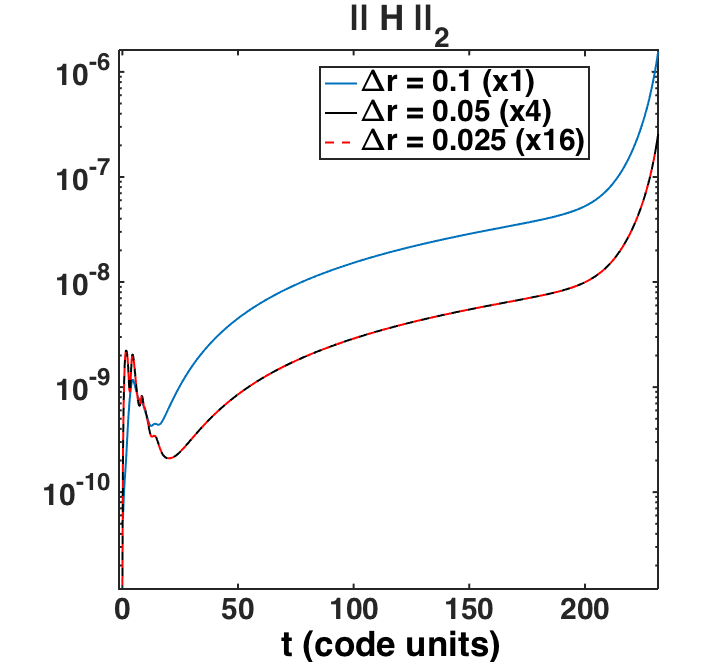}   
    \caption{\footnotesize{Approximation of the $L^2$-norm of the Hamiltonian constraint for simulation of the evolution of a smooth inhomogeneity in the density profile, with three resolutions : $\Delta r = 0.1$, $\Delta r = 0.05$, and $\Delta r = 0.025$. Curve for $\Delta r = 0.05$ has been multiplied by $4$ and curve for $\Delta r = 0.025$ has been multiplied by $16$ to exhibit the second order of convergence of the method.}}
     \label{L2_H}

\end{figure}

\begin{figure}[!]
\centering
	\includegraphics[height = 7cm]{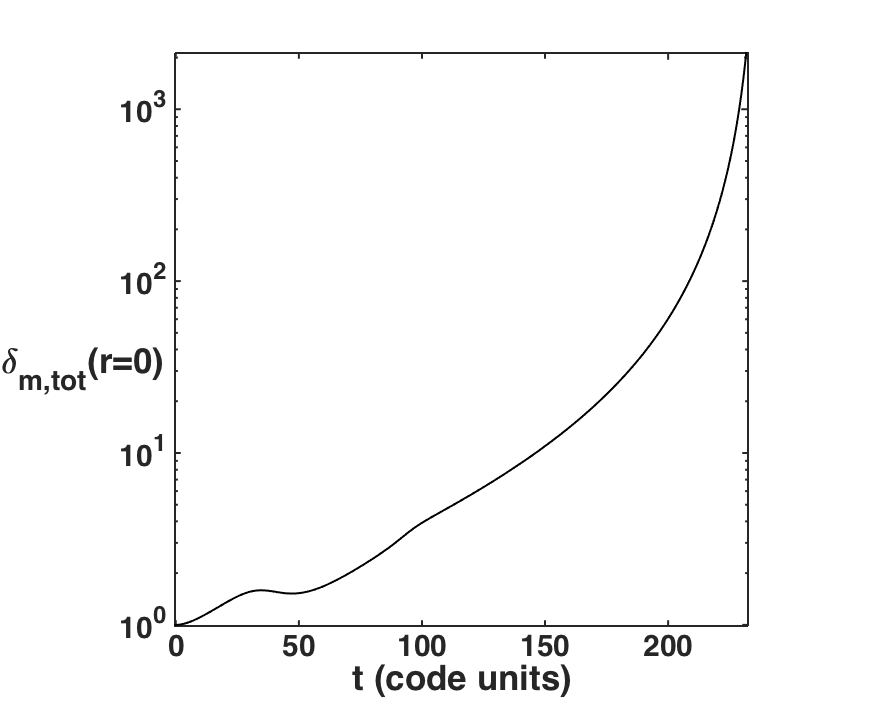}  
    \caption{\footnotesize{Central total energy-density contrast versus time. The divergence indicates a collapse.}}
    \label{delta_e_m_tot}

\end{figure}

Moreover, by inspecting the $L^2$-norm of the Hamiltonian constraint with respect to time, in Fig. \ref{L2_H}, we see that we also have a second order rescaling (as in the previous plot, curve for $\Delta r = 0.05$ has been multiplied by $4$ and curve for $\Delta r = 0.025$ has been multiplied by $16$). The convergence of the method is thus at least second order. Note that the late but steep increase at the end of the simulation is due to the collapse and possibly the reaching of a singularity. Indeed, we can see on Fig. \ref{delta_e_m_tot} that the total energy-density contrast (defined by $\delta_{tot} = \frac{e_{1}+e_{2}}{{\overline{e}_{1}}+{\overline{e}_{2}}}-1$, which is different from $\delta_{1}+\delta_{2}$) seems to diverge. However, note that this is not a significant evidence that a black hole is formed. We will discuss this in the next section.


We thus have validated our integration code since the Hamiltonian constraint has the correct behaviour. We are allowed to trust our simulations and are ready to investigate in the following sections the non linear cosmological spherical collapse with this tool.

\subsection{Typical behaviours and dependence on the equation of state}
\label{section_behaviour}
Before looking for universality with respect to the equation of state, we study the different behaviours of a fluctuation of a single fluid and the influence of the equation of state on it. In all the runs of our code, we fix the following parameters (except if especially mentioned) : $H_i = 0.03$, $a_i = 1$ and $H_0=0.001$.

We perform simulations involving a single barotropic fluid of matter $p = \omega e$ with an initial profile described by the equation \eqref{initial_profile}, which is, as already mentioned, parametrized by three real numbers: the initial amplitude $\delta_i$, the initial size $r_i$ of the fluctuation and the sharpness of the profile $\sigma$.
\begin{figure}[!]
\centering
	\includegraphics[height=7cm]{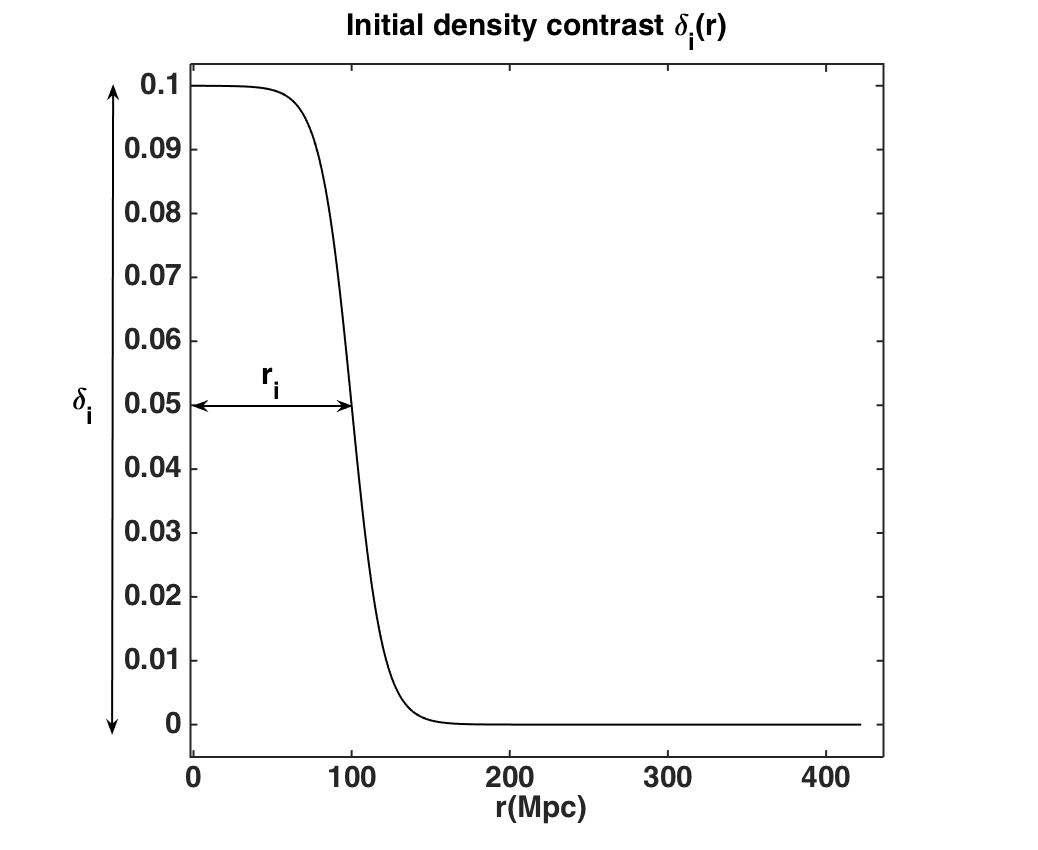} 
	\includegraphics[height=7cm]{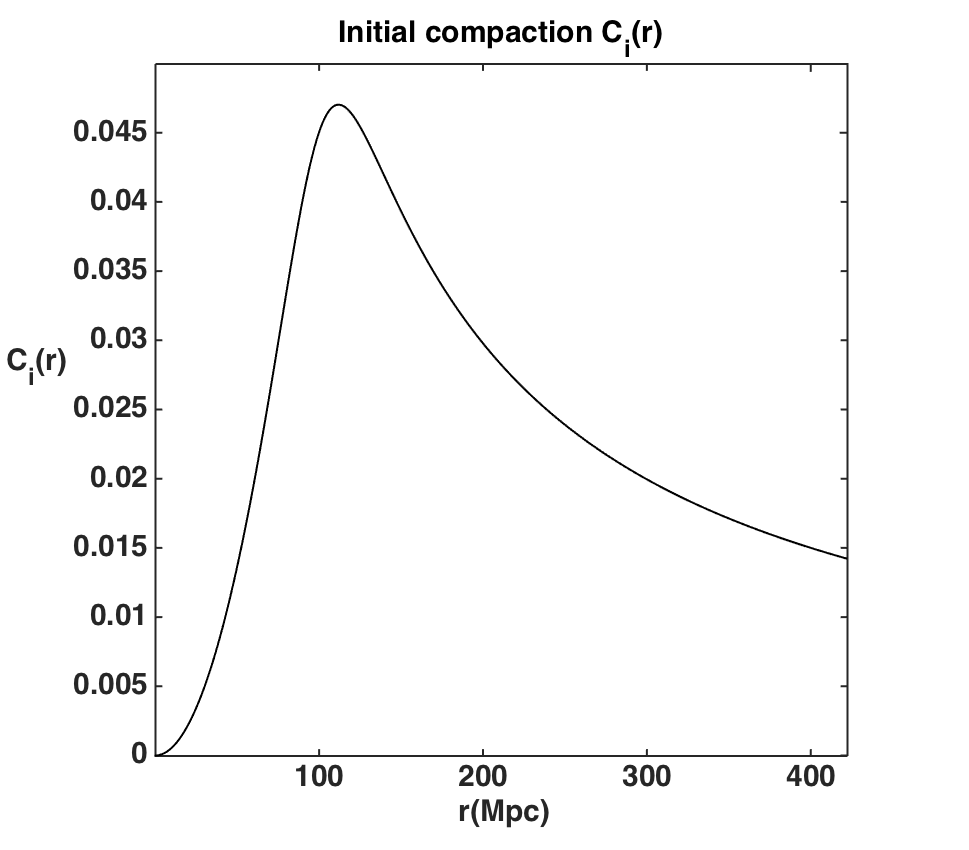}
    \caption{\footnotesize{The upper graph shows the shape of an initial profile of the energy-density contrast. The second one gives the corresponding compaction function. The peak in the latter is in agreement with the size of the fluctuation seen in the above panel.}}
    \label{initial_compact}
\end{figure}
This profile and the corresponding initial compaction function are shown in Fig. \ref{initial_compact}. A compaction function has usually this bell shape: starting from zero, growing to a peak and then decreasing to zero as asymptotic behaviour. The peak determines, as defined in section \ref{used_quantities}, the size of the fluctuation. We can see that it is nearly the same value as $r_i$, which confirms the pertinence of this definition.

Depending on the three initial parameters, $\delta_i$, $r_i$ and $\omega$ ($\sigma$ is fixed to $10\rm Mpc$ in all what follows), we observe two different behaviours. The first one is the collapse while the second one is dilution.

\subsubsection{The collapse scenario}

\begin{figure}
  \centering
  \includegraphics[height=7cm]{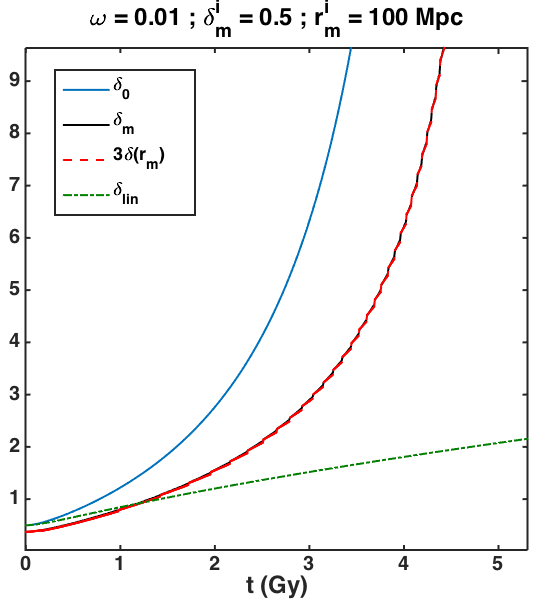}
  \caption{\footnotesize{Evolution of the central energy-density contrast $\delta_0$, the mean energy-density contrast $\overline{\delta}_m$ and the central energy-density contrast $\delta_\text{lin}$ computed with the linear perturbation theory in the collapse scenario. The last curve, $3\delta(r_m)$, shows the validity of the formula \eqref{relation_musco}, although we are not in a super-horizon regime. The full relativist $\delta_0$ and the approximate $\delta_\text{lin}$ are in adequation at early times, which is an additional validation of the code.} \label{deltas_collapse}}\ \\
  \end{figure}
\begin{figure}
  \centering
  \includegraphics[height=7cm]{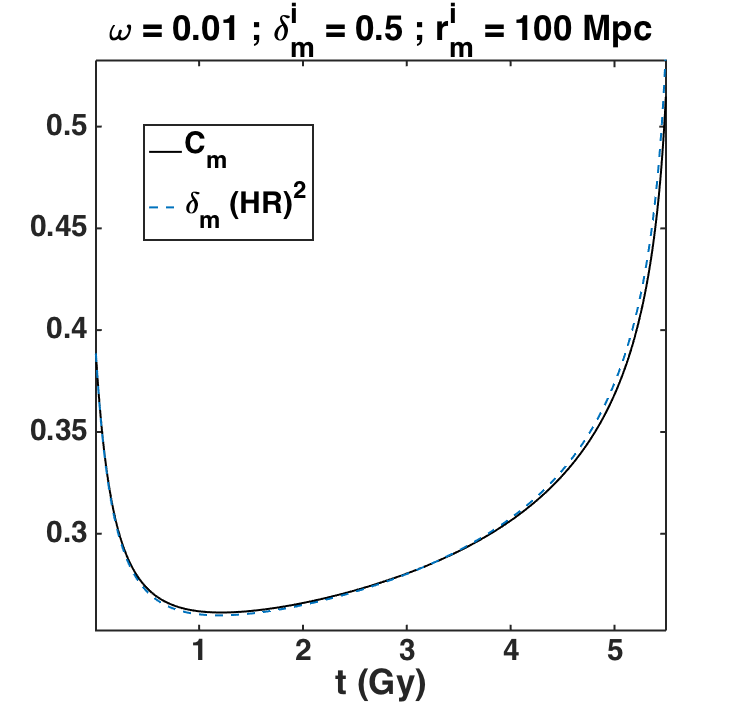}
  \caption{\footnotesize{Evolution of the compactness of the fluctuation. The second curve represents the quantity $\overline{\delta}_m (HR)^2$ and its adequation with $C_m$ confirms the approximation \eqref{relation_harada}, although we are not in a super-horizon regime. The compactness decreases first with the background expansion but then increases more and more rapidly with the collapse. The code is not able to follow it until the black hole formation at $C_m = 1$..} \label{compact_collapse}}
\end{figure}
\begin{figure}
  \centering
  \includegraphics[height=7cm]{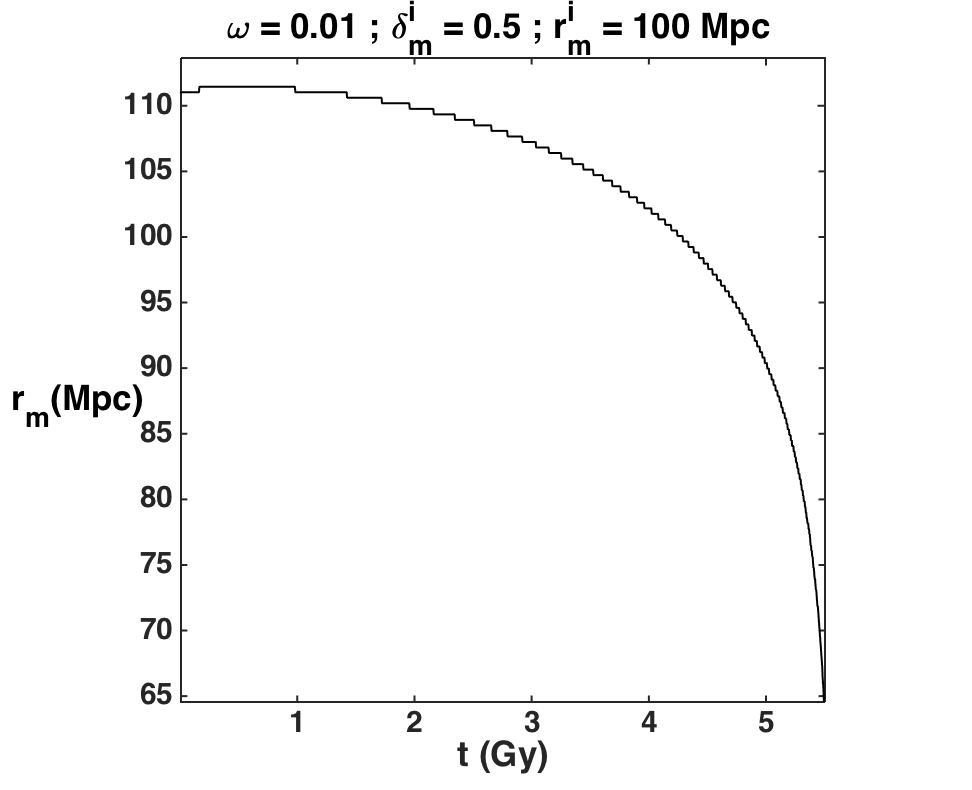}
  \caption{\footnotesize{Evolution of the radius of the fluctuation. The steps come from the spatial discretisation. The radius first increases with the background but then starts decreasing, indicating a concentration of the matter towards the center of the grid.} \label{radius_collapse}}\footnotesize{\ \\
  }
 
 \end{figure}
 \begin{figure}
 \centering
 \includegraphics[height=7cm]{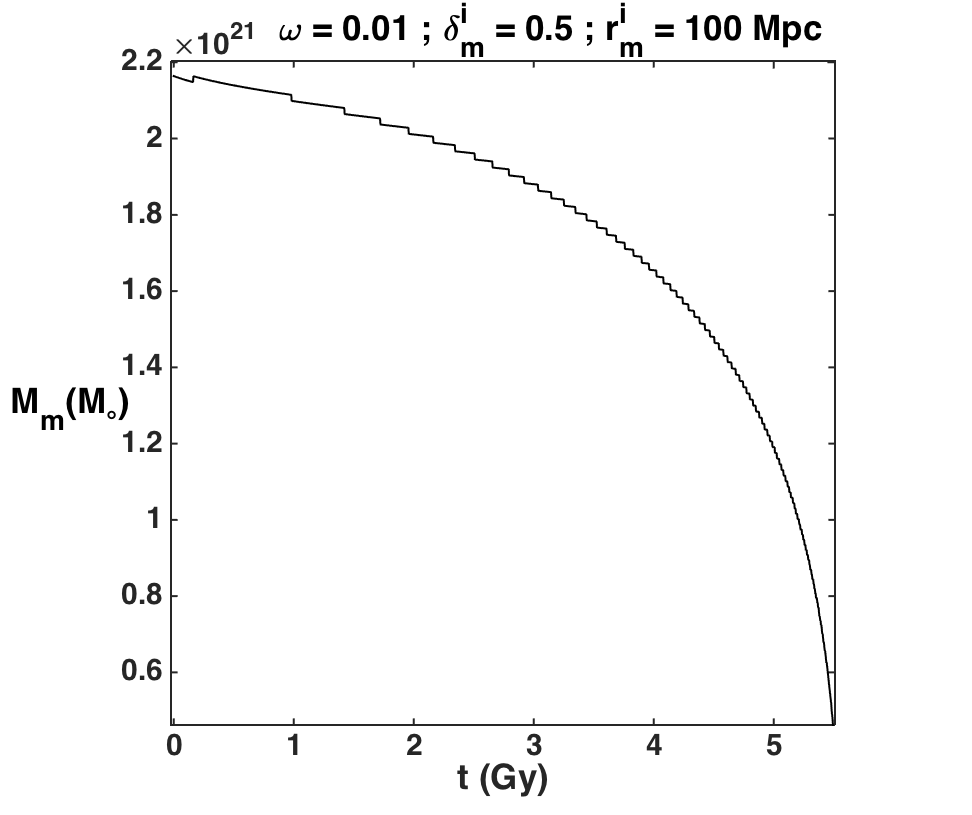}
  \caption{\footnotesize{Evolution of the mass of the fluctuation, in solar mass units. The mass decreases because the integration upper bound is related to the radius of the fluctuation, which is collapsing to zero.} \label{mass_collapse}}\ \\
\end{figure}

We show such an example of collapsing solution in Fig. \ref{deltas_collapse}, \ref{compact_collapse}, \ref{radius_collapse} and \ref{mass_collapse} with the values $\omega = 0.01$, $\delta_i = 0.5$ and $r_i = 100\rm Mpc$. We see on Fig. \ref{deltas_collapse} that the central and mean energy-density contrast are both diverging on the first plot. We also see that the relation \eqref{relation_musco} seems to be correct with good accuracy since the curves of $\overline{\delta}_m$ and $3\delta(r_m)$ are nearly the same, although the conditions for the long-wavelength approximation are not verified in this situation. The last curve represents $\delta_\text{lin}$, the central energy-density contrast computed with the linear perturbation theory (see \cite{Padmanabhan:1993aa} for the basic equations). The agreement between this curve and $\delta_0 := \delta(r=0)$ at early times is an additional indication of the validity of the code. 

Concerning the compactness (Fig. \ref{compact_collapse}), it is first decreasing because of the background expansion. But then it grows until the end of the simulation, indicating a collapse. Our code is not able to follow it at higher compactnesses, but this is sufficient for our purpose. Note that the relation \eqref{relation_harada} is verified all along the collapse, even if we are not in a subhorizon regime. The reason why this relation and \eqref{relation_musco} are verified must probably due to the low value of the equation of state parameter $\omega$. Recall that these relations are perfectly verified in the comoving gauge. And the absence of pressure precisely generates a comoving gauge. Thus the low pressure explains why these relations seem to hold in this case. The Fig. \ref{radius_collapse} shows that the radius of the fluctuation has an increasing phase, coherently with the decreasing phase of the compactness, followed by a fast decreasing. The matter is concentrating towards the center of the grid, which is intuitively logical in the collapse scenario. The Fig. \ref{mass_collapse} shows a decreasing mass. Although this can seem to be illogical, this is normal because the mass we used is an integral whose upper bound is $r_m$, which is decreasing. The steps visible in these two graphs are simply due to the spatial discretization of the grid. Indeed, the radius $r_m$, i.e. the location of the maximal compaction, can only be taken among the spatial values given by the discretization, explaining the discontinuous bumps revealing a change in this value.

\subsubsection{The dilution scenario}

\begin{figure}
  \centering
  \includegraphics[height=7cm]{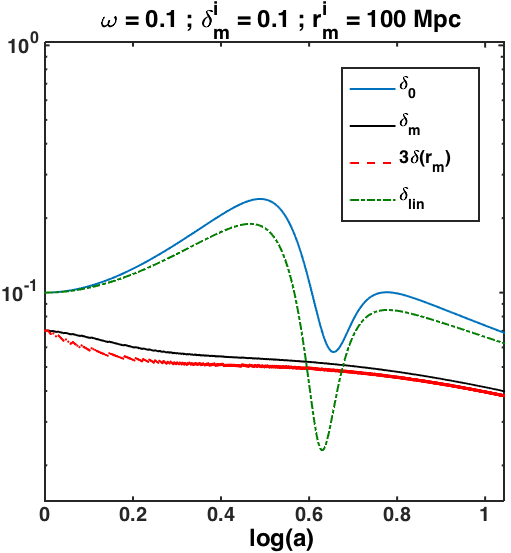}
  \caption{\footnotesize{Evolution of the central energy-density contrast $\delta_0$, the mean energy-density contrast $\overline{\delta}_m$ and the central energy-density contrast $\delta_\text{lin}$ computed with the linear perturbation theory in the dilution scenario. The last curve, $3\delta(r_m)$, shows the inaccuracy of the formula \eqref{relation_musco} in this case. The full relativist $\delta_0$ and the approximate $\delta_\text{lin}$ are in adequation at early times. The full relativistic solution is shown to be in adequation with the linear solution. This confirms the validity of the numerical computation.}  \label{logloga_deltas_new}}
\end{figure}
\begin{figure}
  \centering
  \includegraphics[height=7cm]{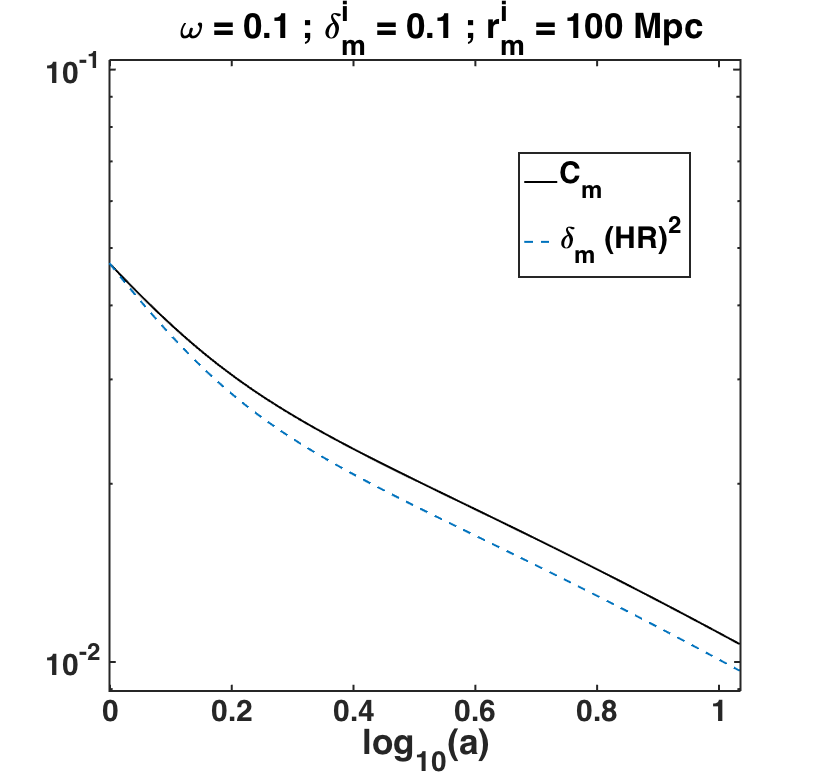}
  \caption{\footnotesize{Evolution of the compactness of the fluctuation. The second curve represents the quantity $\overline{\delta}_m (HR)^2$ and is no more in adequation with $C_m$. The compactness decreases in a powerlaw of the background scale factor, which means that the fluctuation follows its expansion.}  \label{loglog_C_m_vs_a}}
\end{figure}
\begin{figure}
  \centering
  \includegraphics[height=7cm]{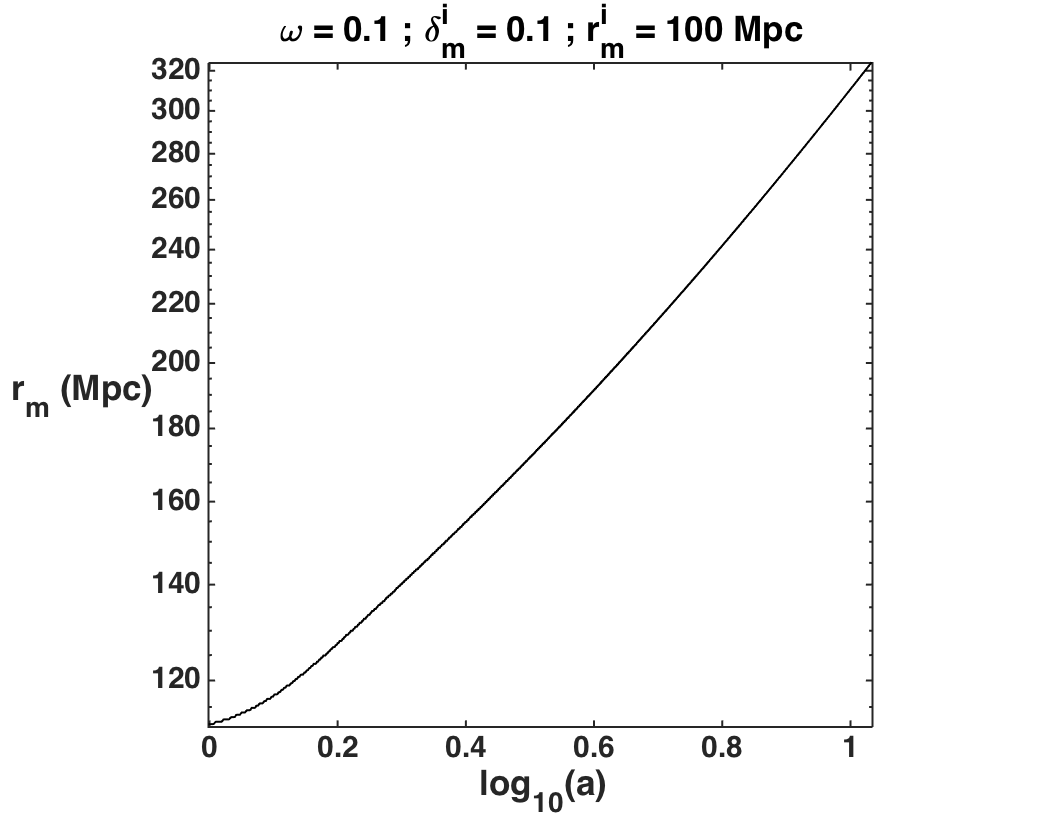}
  \caption{\footnotesize{Evolution of the radius of the fluctuation. It is increasing in a powerlaw of the background scale factor, indicating that the fluctuation follows the cosmological expansion.} \label{loglog_radius_vs_a}}\footnotesize{\ \\
  \ \\
  }
\end{figure}

We give now an example of diluting solution in Fig. \ref{logloga_deltas_new}, \ref{loglog_C_m_vs_a} and \ref{loglog_radius_vs_a} with the values $\omega = 0.1$, $\delta_i = 0.1$ and $r_i = 100\rm Mpc$. First notice that the two correspondences \eqref{relation_musco} and \eqref{relation_harada} are no more exactly verified because we involved more pressure and the long-wavelength approximation does not hold. Then, on the first plot, the central energy-density contrast has an oscillations phase before decreasing to zero in a power law of $a$. This is in agreement with the linear perturbation theory, which gives another validation of the numerical computations. The mean energy-density is also decreasing but without oscillations.
Fig. \ref{loglog_C_m_vs_a} and \ref{loglog_radius_vs_a} show, through clear power laws in $a$, that the fluctuation is at late time completely diluted in the background and only follows the dynamics of the latter. The background expansion and the internal pressure are too strong for the fluctuation to collapse and make it disappear.\\

\subsection{The $1+\log$ slicing to simulate black holes formation}
\label{sec:up_and_down_simulation}

We now test another slicing condition, the $1+\log$ slicing presented in section \ref{slicing_conditions}, to try to follow the collapse until the formation of a black hole. With this slicing, we give here an example of a sub and a super-critical solution obtained in that gauge.

\begin{figure}[!]
\centering
	\includegraphics[height=7cm]{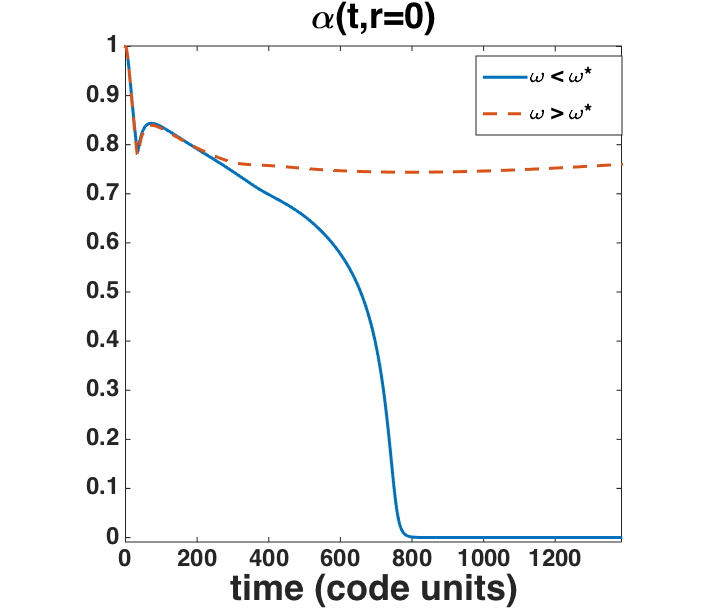} 
    \caption{\footnotesize{Central value of the laps in a sub and a super-critical solution. The sub-critical solution leads to the formation of a black hole as the collapse of the lapse indicates it. The super-critical solution shows a lapse regrowing indicating a dilution.}}
    \label{alpha_up_and_down}
\end{figure}

The central value of the lapse is shown in Fig. \ref{alpha_up_and_down}. The sub-critical solution is a collapsing solution and we can see the formation of a black hole since the lapse is collapsing to zero. This freezing of time at the centre permits the code to avoid the central singularity and to pursue the calculation outside the horizon after its formation. On the contrary, the other solution exhibits a lapse regrowing away from zero, indicating that the overdensity is diluting. At early times, both solutions follow nearly the same curve before choosing between a collapsing or a diluting solution. This is typically what occurs in critical phenomena : near-critical solution all follow the critical (self-similar) solution until falling into a black hole solution or a diluting one. The time before leaving this curve depends on the closeness to the critical parameter $\omega^*$.

\begin{figure}[!]
\centering
	\includegraphics[height=7cm]{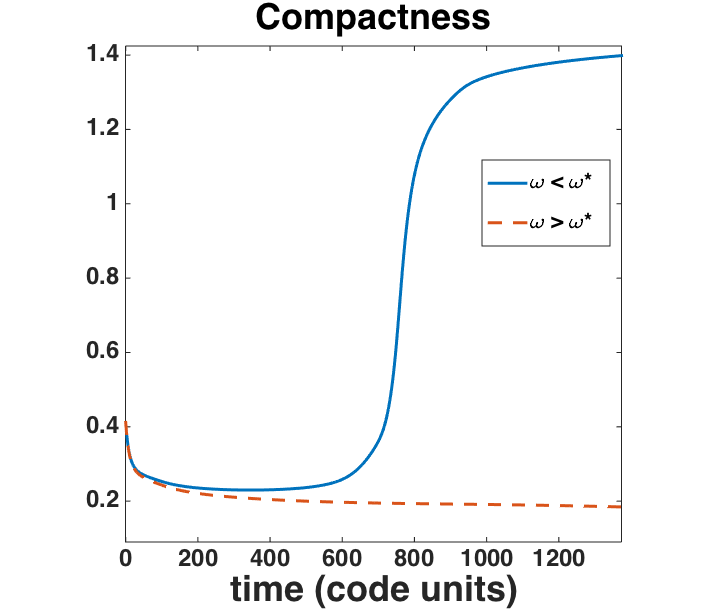} 
    \caption{\footnotesize{Evolution of the compactness in a sub and a super-critical solution. The compactness reaches values greater than unity for the sub-critical solution, revealing the formation of a black hole. Its final compactness seems to be around $1.4$. The diluting super-critical solution shows a decreasing compactness.}}
    \label{compactness_up_and_down}
\end{figure}

The compactnesses are presented in Fig. \ref{compactness_up_and_down}. The collapsing solution shows a compactness growing higher than unity, the black hole limit. But the final compactness of the object seems to stabilise to a value near $1.4$. On the other hand, in the diluting solution, the compactness decreases to zero, indicating that the over-density is progressively disappearing.

\begin{figure}[!]
\centering
	\includegraphics[height=7cm]{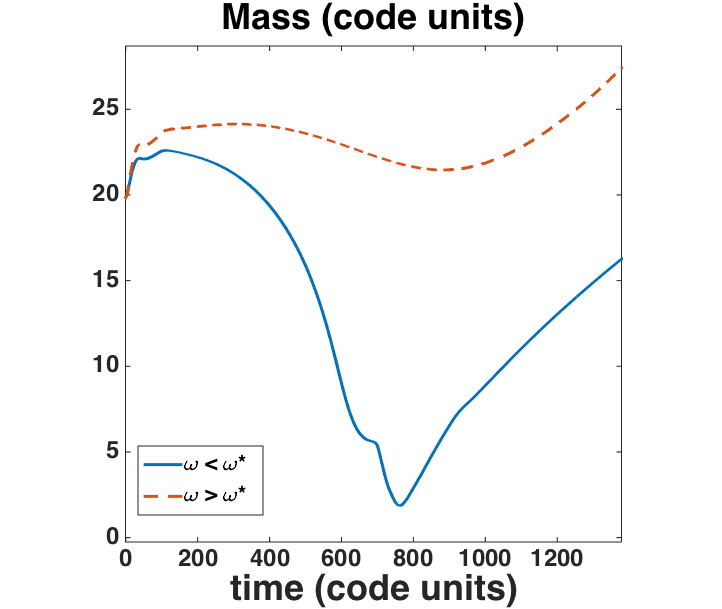} 
    \caption{\footnotesize{Evolution of the mass in a sub and a super-critical solution. In the collapsing solution, the mass is decreasing until the black hole formation. After that, it regrows, showing that the formed black hole attracts matter. The other solution shows equally a growing mass because of the radius of the fluctuation grows with the background expansion.}}
    \label{mass_up_and_down}
\end{figure}

Concerning the mass, we see in Fig. \ref{mass_up_and_down} that it is decreasing until the formation of the black hole. Then, once it has formed, the black hole grows and accretes matter. The diluting solution shows a late-time evolution following the cosmological expansion and thus a late-time growing mass because of the growing of $r_m$.

\begin{figure}[!]
\centering
	\includegraphics[height=7cm]{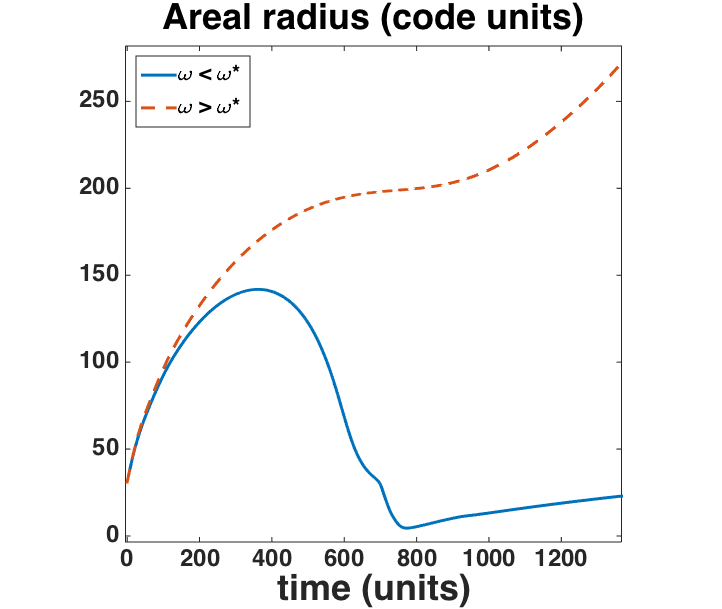} 
    \caption{\footnotesize{Evolution of the areal radius in a sub and a super-critical solution. The collapsing solution shows a size that is decreasing until the formation of a black hole and after that is regrowing.}}
    \label{areal_radius_up_and_down}
\end{figure}

The shape of the areal radius, in Fig. \ref{areal_radius_up_and_down}, equally shows that the size of the black hole regrows once it has formed, due to the matter accretion, and that, in the diluting solution, it is always growing.

With all these information, the question is thus "Which value for the mass of the formed black hole should we take?". As we saw, it is not stabilizing to a constant value but is growing after the black hole formation. Thus, the most reasonable choice is to take the value of the mass when the compactness is equal to one, which corresponds almost to the minimum in the curves of the mass and of the areal radius. 

\subsection{Universality in the critical phenomenon}
\label{universality}

We now want to see if the spherical collapse is universal with respect to matter species. For this reason, we must check if the relation \eqref{universal_powerlaw} is verified if we take for the parameter $k$ the equation of state parameter $\omega$ (although this parameter does not represent exactly what is called strictly speaking the initial conditions), that is to say :
\begin{equation}
M \propto | \omega - \omega^* |^\gamma
\label{universality_vs_omega}
\end{equation}
with $\gamma$ a constant independent of $\omega$. We test this hypothesis in three different cosmologies : the empty Minkowski space-time, the empty de Sitter space-time and the full of matter Einstein-de Sitter space-time.

\subsubsection{The Minkowski case}

This case is, by far, the easiest one. Our code was not built to deal with an empty background but the only differences consist in the scales and the initial conditions. The time, length and mass scales are, here, given by $t_\text{scale} = \frac{GM_\odot}{c^3}$ (in $\text{s}$), $l_\text{scale} = \frac{GM_\odot}{c^2}$ (in $\text{m}$) and $m_\text{scale} = M_\odot$ (in $\text{kg}$), while the initial profile is based on the energy-density profile instead of the energy-density contrast profile because $\overline{e}_m = 0$ :
\begin{equation}
e^i_{m} (r) = e^i_{m} \frac{1-\tanh\left(\frac{r-{r^i_{m}}}{2} \right) }{1+\tanh\left(\frac{r^i_{m}}{2}\right)},
\label{initial_profile_mink}
\end{equation}
where $e^i_m$ is the initial amplitude of the object and $r^i_m$ its initial radius. We work in code units and take as initial conditions $e^i_m = 10^{-5}$ (which corresponds to $6.18\times 10^{15}$ $\frac{\rm kg}{\rm m^3}$) and $r^i_m = 20$ (which corresponds to $2.95 \times 10^4$ $\rm m$), with an initial compactness of $0.048$. We use the $1+\log$ slicing to observe the complete formation of black holes.
\begin{figure}[!]
\centering
	\includegraphics[height=7cm]{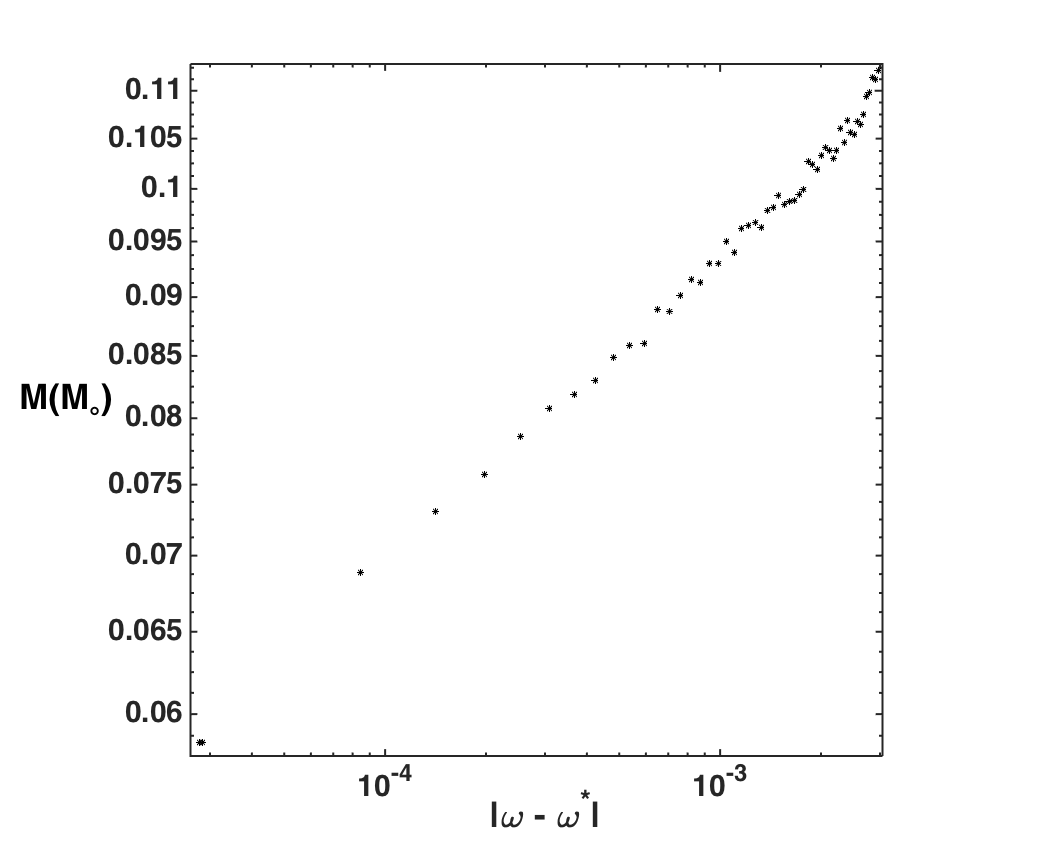} 
    \caption{\footnotesize{Mass of the formed black hole in a Minkowski background computed with the $1+\log$ slicing. The power law is clearly visible, proving universality in this case.}}
    \label{power_law_q_mink}
\end{figure}
All this gives us as critical $\omega^*$ the value of $0.0094$ and the evolution of the mass of the black hole, that we follow until its formation, with respect to $| \omega - \omega^*|$ is shown in Fig. \ref{power_law_q_mink}. In this plot, we observe that all points lie nearly perfectly along a straight line, indicating a power law and thus universality. This is in agreement with \cite{Choptuik:1993aa} and generalises universality to one particular 1-parameter family of matter species, in a Minkowski background.

\subsubsection{The de Sitter case}

Our second test will be the case of an empty space with a positive cosmological constant $\Lambda$. We use the same energy profile as in the previous test with $e^i_m = 10^{-5}$ and  $r^i_m = 20$. The initial Hubble factor is set to $H_i = 3\times 10^{-5}$ and $\Omega_\Lambda = 1$, in such a way that the value of $\Lambda$ is fixed by $H_i$. The $1+\log$ slicing is used to follow the black holes formations. 

\begin{figure}[!]
\centering
	\includegraphics[height=7cm]{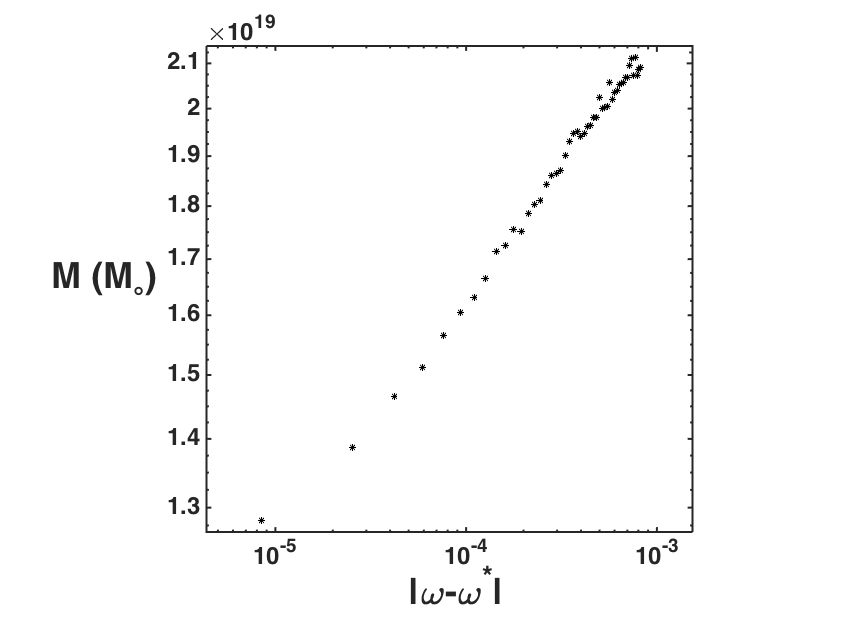} 
    \caption{\footnotesize{Mass of the formed black hole in a de Sitter background computed with the $1+\log$ slicing. The power law is clearly visible, proving universality in this case.}}
    \label{powerlaw_de_sitter}
\end{figure}

With this, the critical solution appears to be around $\omega^* \simeq 0.0128$. The mass of the formed black holes, with respect to $|\omega-\omega^*|$ is shown in Fig. \ref{powerlaw_de_sitter}. We observe that all the points are along a straight line, revealing the universal scaling law. As well as in the Minkowski case, we thus proved universality in the de Sitter case which is relevant from the cosmological point of view. 

\subsubsection{The Einstein-de Sitter case}

The Einstein-de Sitter case is more complicated because of the presence of matter at the outer boundary. This non zero asymptotic value for the energy-density renders the code less stable than in the two previous cases. However, this code is able to follow the formation of a black holes even in this case. We will just not be able to go as close to the critical solution as we did for the Minkowski and the de Sitter cases. Moreover, the universal scaling law differs a bit from these two cases because it is not the mass of the black hole that is rescaling but the ratio $\frac{M}{M_H}$ where $M_H$ is the horizon mass computed at the time when the areal radius of the fluctuation is equal to the Hubble radius. We thus want to prove that 
\begin{equation}
\frac{M}{M_H} \propto | \omega - \omega^* |^\gamma
\label{universality_vs_omega2}
\end{equation}
with $\gamma$ a constant independent of $\omega$.

To achieve that, we first test the conformally flat initial conditions of Sec. \ref{conformally_flat} with the smooth top-hat profile \eqref{initial_profile}. Values taken for the profile are $\delta^i_m = 0.5$,  $r_i = 100\rm Mpc$ and $\sigma = 10\rm Mpc$ and gives $\omega^* = 0.058$ as a critical value.

\begin{figure}[!]
\centering
	\includegraphics[height=7cm]{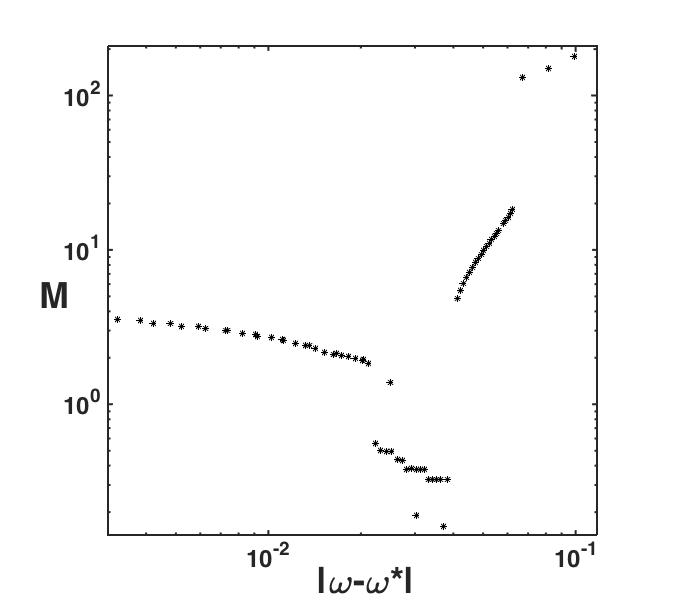} 
    \caption{\footnotesize{Mass, by units of horizon masses, of the formed black hole computed with the $1+\log$ slicing. No scaling law is observed, indicating that values closer to $\omega^* = 0.058$ should be computed to look for universality.}}
    \label{mass_scaling_1log}
\end{figure}

The mass values computed in this way are reported on Fig. \ref{mass_scaling_1log}. Unfortunately, the results are not in favour of universality : no scaling law is visible and, worse, no decreasing of the mass is seen when considering the most left points in the graph. We are thus in a similar case to what is described in \citep{Hawke:2002aa}. In \cite{Musco:2009aa}, it is explained that this behaviour appears because taking non linear initial profiles (with too large initial amplitude $\delta_i$)  whose length-scale are smaller than the cosmological horizon generates shocks in near critical solutions because of the presence of decreasing modes. We will thus need to test super-horizon fluctuations to try and avoid this behaviour.

We thus follow the advises of \cite{Musco:2009aa} and take a fluctuation starting in the super-horizon regime that admits only growing modes when entering the horizon. To achieve that, we use initial conditions computed from the long-wavelength approximation presented in section \ref{sec:long_walength}, still in the $1+\log$ slicing. The parameter $\Delta$ of profile \eqref{profile_zeta} is chosen such that the initial areal radius of the fluctuation is three-times the Hubble radius $R_H = H^{-1}$. In such a way, all the decreasing modes have disappeared when the size of the inhomogeneity coincides with the cosmological horizon. The second parameter, $A$, denotes the amplitude of the overdensity and will thus also influence the value of the critical equation of state parameter $\omega^*$. We perform two sets of simulation, one with $\omega_1^* \simeq 0.337$ and the other with $\omega_2^* \simeq 0.434$.

\begin{figure}[!]
\centering
	\includegraphics[height=7cm]{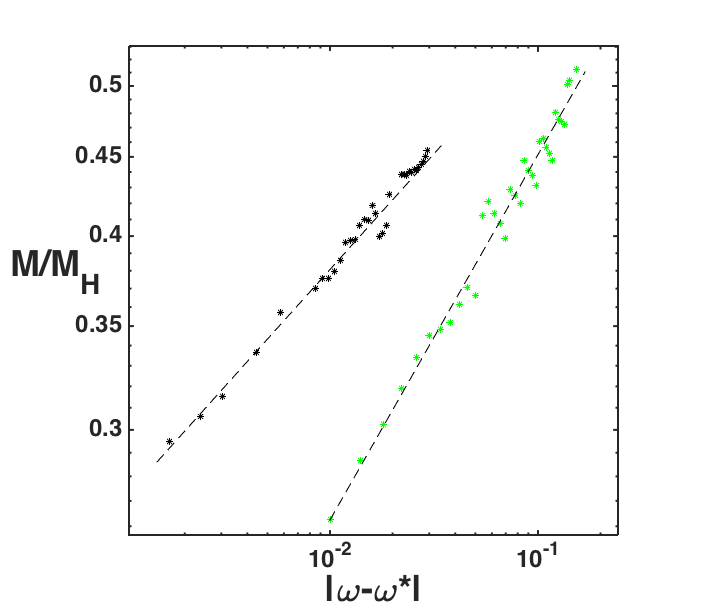} 
    \caption{\footnotesize{Mass, by units of horizon masses, of the formed black hole computed with the $1+\log$ slicing for two families of profiles. Black dots represent family $1$, with $\omega_1^* \simeq 0.334$, while green ones represent family $2$, with $\omega_2^* \simeq 0.434$. In both cases, dots seem to be globally on the same line, indicating a power-law and, thus, that the universality relation is verified. The slope of the line depends on the family of profiles. }}
    \label{powerlaw_lwa2}
\end{figure}

 The masses of the formed black holes, rescaled by the horizon mass $M_H$ (the mass inside the horizon computed at the time the fluctuation enters the Hubble radius), are shown in Fig. \ref{powerlaw_lwa2}. It is clearly visible, in both cases, that the plotted quantity is rescaling in a power-law of $|\omega - \omega^*|$, although our code does not permit to obtain perfect lines and values nearer the critical solution. This is thus a serious indication in favour of universality. Note that the critical exponent $\gamma$ is shown to vary as a function of the amplitude of the profiles. 

 We can draw two major conclusions from this. First, our simulations with sub-horizon and conformally flat initial conditions, though interesting and relevant, revealed a breakdown of universality close to $\omega^*$. Secondly, the universal scaling law seems to be verified when using super-horizon initial conditions with exclusively growing modes. This is perfectly in agreement with \cite{Hawke:2002aa}, \cite{Musco:2009aa}  and \cite{Musco:2013aa}. They worked by fixing the equation of state and varying the matter profile. On our side, we showed universality with respect to the matter type by varying the equation of state parameter and this is the major original result of this work. It should however be checked with a more efficient code, equipped with an adaptive mesh-refinement specially designed for spherical symmetric space-times like the one used by \cite{Musco:2013aa}, to be proven definitely.

So, with all these simulations, using different gauges and backgrounds, we can be confident with our results and conclude that the spherical collapse is fully universal (with respect to the equation of state) in a Minkowski background and partially universal in a full matter background. We end by saying that we guess that the critical exponent of the power law must depend on the initial profile and its value, in itself, should thus be less fundamental than those found by varying the initial conditions for a fixed matter specie.

\section{Conclusion}

In this work we have upgraded the BSSN code used in \cite{Rekier:2015aa} and \cite{Rekier:2016aa} to study the spherical collapse of pressured matter thanks to the addition of a HLLE incomplete Riemann solver for the relativistic hydrodynamics. This code is now able to deal with several matter species, thanks to the use of a non comoving gauge, in a general Friedmann universe. With it, we could focus on fluctuations of a single barotropic fluid and investigate the critical collapse with respect to the constant parameter $\omega = \frac{p}{e}$ of the equation of state. 

In GR, the question of the observables is always tricky because of the gauge dependence of the tensors components. Even such an important quantity as the energy-density contrast is gauge dependent because it consists in a local-background comparison of variables that do not share the same proper time. To avoid difficulties of interpretation, we worked first in the synchronous gauge which was good enough for our preliminary observations despite its well-known instability. But we used the $1+\log$ slicing, which was found more appropriated, to simulate black holes formations in the second part of the work.

Concerning the universality of the critical collapse, we saw that, for a fixed profile of the energy-density contrast, there is a critical value $\omega^*$ of the equation of state parameter under which the fluctuation collapses to a black hole and upper which it is diluting to the background. For sub-critical solutions, that is to say collapsing solutions, we tried to prove numerically the universal scaling law for the mass of the formed black hole $M \propto |\omega - \omega^* |^\gamma $ similar to what is found in well known critical phenomena, with a critical exponent independent on the value of the equation of state parameter $\omega$. This has been shown in the Minkowski and the de Sitter cases. For the Einstein-de Sitter universe, we first saw no scaling law, possibly because we could not go as close as necessary to the critical solution to observe it. The other possibility is that we encountered a similar behaviour as in \cite{Hawke:2002aa} who saw a breaking of universality near the critical solution because of the use of conformally flat and sub-horizon initial conditions that possess decreasing modes. To bypass this problem, we used super-horizon initial fluctuations with exclusively growing modes. We thus derived, in any Bona-Masso slicing given by eq. \eqref{slicing}, the solution of the equations in the long-wavelength approximation. This original development, which is interesting in itself because of its similitude to the same solution in \cite{Harada:2015aa} in other gauges, allowed us to evolve more realistic initial conditions. With it, we could finally observe the universality scaling law we were looking for. The result is not perfect since we could not approach the critical solution as close as we wished, an adaptive mesh refinement should be used to perform this, but it gives more than an indication that universality is true in this case.

In conclusion, we have shown, among others, that the spherical collapse of a barotropic fluid $p=\omega e$ is universal with respect to the value of the equation of state parameter. These results are important because it is a generalisation of numerous previous works: the gravitational spherical collapse appears to happen in a very general way that depends little on the initial curvature profile and, we know it now thanks to our own work, the types of matter (through the equation of state parameter $\omega$). Our treatment of the problem enforces the importance of the use of numerical relativity when dealing with pressured fields of matter. We indeed also exhibited scenarios where the linear approach fails to give an acceptable approximation of the solution, even at small amplitudes. 

The critical collapse and, more generally, the spherical collapse are studies that can be declined in numerous ways by using numerical relativity, depending on the fluid(s) of matter, the Dark Energy model, the cosmological epoch, the scale of interest, the coupling between the source terms,\ldots This article provides lots of perspectives for future works. We give now some of these in what follows.

We used the barotropic linear equation of state $p=\omega e$, where $e$ is the energy-density and $p$ is the pressure because this is what is commonly considered in cosmology. However, other equations of state are often used in the astrophysical case, such as the one for an ideal fluid \eqref{eos_ideal} and the one for a polytropic fluid \eqref{eos_polytropic}. Threshold values of the parameters contained in these equations and the associated critical phenomenon could be examined. This would extend again universality with respect to a larger panel of matter species or, on the contrary, lead to the discovery of a breaking of the universality.

As explained in the Introduction and in section \ref{validation}, the code was designed to evolve possibly two non interacting fluids of matter, with a scalar field and a cosmological constant. In this work, we only explored the case of a single fluid of matter but any other combination of the previous ingredients is possible. For example, the late time evolution of pressured matter minimally (or not) coupled to a scalar field would have interesting cosmology applications. A second obvious application would be the study of structure formation at the epoch of matter-radiation equivalence since we can evolve conjointly dust and radiation, during the inflation era or in a de Sitter background. A third application would be the study of voids since a negative energy-density contrast can easily be implemented.

From the numerical relativity point of view, we built an code which appeared to have some imperfections. Some improvements could be performed, such as the implementation of an adaptive mesh refinement or a logarithmic spatial discretization. But the question of the implementation of the cosmological boundary conditions is more tricky and would probably require a research work in itself. Numerical relativity is a subject currently on the rise, thanks to the active field of gravitational waves. It is becoming more and more employed within the frame of cosmology, especially concerning the PBH formation but not only, and we hope to have enhance a bit the marriage of the two disciplines.\\

\noindent \textit{Acknowledgments}
The authors warmly thanks Dr. I. Cordero-Carrión for helpful advices and discussions concerning the Valencia formulation and the numerical methods used in this work.
F. S. is supported by a FRS-FNRS (Belgian Funds for Scientific Research) Research Fellowship. J.R. is supported by the European Research Council (ERC) under the European Union's Horizon 2020 research and innovation programme (Advanced Grant agreement No. 670874). This research used resources of the "Plateforme Technologique de Calcul Intensif (PTCI)" (http://www.ptci.unamur.be) located at the University of Namur, Belgium, which is supported by the F.R.S.-FNRS under the convention No. 2.5020.11. The PTCI is member of the "Consortium des Équipements de Calcul Intensif (CÉCI)" (http://www.ceci-hpc.be).

\section*{Appendix A : Recovering the primitive variables from the conserved ones}

The conserved variables are defined in such a way : 
\begin{equation}
\begin{cases}
D &= \rho W\\
S_i &= (e+p)W^2v_i\\
\tau &= (e+p)W^2-p-D
\end{cases}
\label{relation_primtocons}
\end{equation}
where the rest-mass density $\rho$, the energy density $e$, the pressure $p$, the velocity $v_i$ and the Lorentz factor $W = \frac{1}{\sqrt{1-v^2}}$ are the primitive variables. In general, the inversion of this relation is not analytical and requires a numerical root-finding procedure (see \cite{Rezzolla:2013aa} for more details).

In the case of the barotropic equation of state $p = \omega e$, this inversion can be made analytically. To obtain it, we start by squaring the second equation of \eqref{relation_primtocons} :
\begin{equation}
S^2 := S^iS_i = (e+p)^2W^4v^2.
\label{Ssquare1}
\end{equation}
Recalling that $v^2 = 1-\frac{1}{W^2}$, \eqref{Ssquare1} becomes
\begin{equation}
S^2 = (e+p)^2W^4 - (e+p)^2W^2.
\label{Ssquare2}
\end{equation}
The third equation of \eqref{relation_primtocons} gives
\begin{eqnarray*}
(e+p)^2W^4 &=& (\tau +D+p)^2\\
(e+p)^2W^2 &=& (e+p)(\tau +D+p). 
\end{eqnarray*}
Reinserting in \eqref{Ssquare2} gives the relation\footnote{Note that this relation is correct whatever the equation of state and does not requires spherical symmetry.}
\begin{equation}
S^2 = (\tau +D)^2  + (\tau + D)(p-e) - pe.
\end{equation}
Using now the equation of state $p=\omega e$, we obtain a quadratic equation in $e$ : 
\begin{equation}
-\omega e^2 + (\omega-1)(\tau +D)e + (\tau +D)^2 - S^2 = 0.
\end{equation}
Its solutions are 
\begin{equation}
e = (\tau +D) - \frac{S^2}{\tau +D}, \text{\ \ \ \ if $\omega=0$},
\end{equation}
and
\begin{equation}
e = \tau +D, \text{\ \ \ \ if $\omega=-1$},
\end{equation}
and
\begin{eqnarray}
e_\pm &=& \frac{(\omega-1)(\tau +D) \pm \sqrt{(\omega +1 ) (\tau +D)^2 -4\omega S^2}}{2\omega},\nonumber\\
 &&\text{\ \ \ \ if $\omega\neq 0$}
\end{eqnarray}
The sign we have to consider depends on the value of $\omega$. 

If $0 < \omega \leq 1$, we have $(\omega - 1)(\tau +D) \leq 0$ and thus we have to take the plus to keep a non negative energy density. 

If $\omega > 1$, the positivity of the interior of the root and the fact that 
\begin{equation}
(\omega +1)(\tau +D)^2 - 4\omega S^2 \leq (\omega +1)\left[(\tau +D)^2-S^2\right]
\end{equation}
imply that 
\begin{equation}
e_+e_- = \frac{(\tau +D)^2-S^2}{-\omega} \leq 0,
\end{equation}
where the equality holds if and only if $e = 0$. The latter case is trivial,because it requires $D = 0$, $S_i = 0$ and $\tau = 0$, and in the other cases, this means that the solutions $e_+$ and $e_-$ have opposite signs and that only one is positive.

If $\omega < 0$, the situation is less obvious because both solutions can be positive. For example, if $\omega \in ]-1;0[$, we have 
\begin{eqnarray*}
e_+e_- &=& \frac{(\tau +D)^2-S^2}{-\omega}\\
 &=&\frac{\left((e+p)W^2-p\right)^2-(e+p)^2W^4v^2}{-\omega} \\
&=& \frac{\left((\omega +1)^2W^4 +\omega^2 - 2 \omega (\omega +1)W^2\right)e^2}{-\omega} \\
&&-\frac{ (\omega + 1)^2W^4v^2e^2 }{-\omega}\\
&=& \frac{\left[ (\omega +1)^2W^4(1-v^2) + \omega^2 - 2\omega (\omega +1)W^2\right]e^2 }{-\omega}\\
&>& 0,
\end{eqnarray*}
because all the terms of the numerator are positive. This means that both solutions have the same sign and thus are positive. The choice must be done thanks to the continuity of the solution with time, which can be difficult numerically.

Ones the energy density $e$ is computed, the other variables follow easily :
\begin{eqnarray}
p &=& \omega e\\
W &=& \sqrt{\frac{\tau +D +p}{e+p}}\\
v_i &=& \frac{S_i}{(e+p)W^2}\\
\rho &=& \frac{D}{W}.
\end{eqnarray}

\section*{Appendix B : Splitting for the source terms in the PIRK operators}

The evolution equations are written in the form \eqref{PIRK operators} because we are using the PIRK algorithm. The splitting has been chosen to ensure the scheme to be as stable as possible (see \cite{Rekier:2015aa}). In the first step, the hydrodynamical conserved variables, the cosmological scale factor $a$, the lapse $\alpha$, the elements of the conformal $3$-metric $\hat{a}$ and $\hat{b}$ and $\psi$ are evolved explicitly. These are thus included in the $L_1$ operator. In the second step, the extrinsic curvature is evolved. This means that $K$ and $A_a$ are split into the following $L_2$ and $L_3$ operators \footnote{The terms of $R^r_r$ and $R$ which are proportional to $\hat{\Delta}^r$ and $\partial_r \hat{\Delta}^r$ are in fact included in the $L_{3(A_a)}$ operator instead of $L_{2(A_a)}$.} :
\begin{eqnarray}
L_{2(A_a)} &=& -\left(\nabla^r\nabla_r \alpha - \frac{1}{3}\nabla^2 \alpha \right) \nonumber\\
&&+ \alpha\left( R^r_r - \frac{1}{3}R\right), \\
L_{3(A_a)} &=& \alpha K A_a - \frac{16\pi}{3}\left(S_a - S_b\right),\\
L_{2(K)} &=& -\nabla^2\alpha, \\
L_{3(K)} &=& \alpha \left( A_a^2 + 2A_b^2 + \frac{1}{3} K^2 \right)\nonumber\\
&& + 4\pi \alpha \left( E + S_a + 2S_b\right).
\end{eqnarray}
Finally, the auxiliary variable $\hat{\Delta}^r$ is evolved partially implicitly : 
\begin{eqnarray}
L_{2\left(\hat{\Delta}^r\right)} &=& -\frac{2}{\hat{a}}\left(A_a\partial_r \alpha + \alpha\partial_r A_a\right)  - \frac{4\alpha}{r\hat{b}}\left(A_a -A_b\right) \nonumber\\
 && + \frac{\xi\alpha}{\hat{a}}\Bigg[\partial_r A_a - \frac{2}{3}\partial_r K + 6A_a\frac{\partial_r\psi}{\psi}\nonumber \\
 &&+ \left(A_a-A_b\right)\left(\frac{2}{r}+\frac{\partial_r\hat{b}}{\hat{b}}\right) \Bigg], \\
 L_{3\left(\hat{\Delta}^r\right)} &=& 2\alpha A_a \hat{\Delta}^r - 8\pi j_r \frac{\xi\alpha}{\hat{a}}.
\end{eqnarray}
Note that general expressions can be found in \cite{Montero:2012aa}.

\section*{Appendix C : Kodama mass and mean energy-density contrast in BSSN variables}

The Kodama mass was first defined in \cite{Kodama:1980aa} but we take \cite{Shibata:1999aa} and \cite{Harada:2015aa} as references.

Recall that, in spherical symmetry, the areal radius is the positive quantity $R(t,r)$ defined by the area $A(t,r)$ of the surface defined by constant $t$ and $r$ coordinates in such a way :
\begin{equation}
A = 4\pi R^2.
\end{equation}
In our BSSN metric \eqref{metric}, the areal radius is simply the square root of its $\theta\theta$ component :
\begin{equation}
R = \sqrt{g_{\theta\theta}} = \psi^2a\sqrt{\hat{b}}r.
\end{equation}
Consider now the $2$-metric $G_{AB} = \begin{pmatrix}
g_{tt} & g_{tr}\\
g_{rt} & g_{rr}
\end{pmatrix}$ with $A,B\in {t,r}$. We define the Kodama vector by 
\begin{equation}
K^A = \epsilon^{AB}\partial_BR,
\end{equation}
where $\epsilon_{AB} = \sqrt{-G}\varepsilon_{AB}$ with $\varepsilon_{AB}$ being the Levi-Civita symbol and $\epsilon^{AB} = G^{AC}G^{BD}\epsilon_{CD}$. Working in the zero shift gauge gives
\begin{eqnarray}
K^t  &=& -\frac{\partial_rR}{\alpha \psi^2a\sqrt{\hat{a}}} \\
K^r &=& \frac{\partial_tR}{\alpha\psi^2a\sqrt{\hat{a}}}.
\end{eqnarray}
The tensor $K^A$ is extended to a $4$-vector $K^\mu$ by posing $K^\theta = K^\phi = 0$. The quantity $S^\mu = T^\mu_\nu K^\nu$ is thus a conserved current (see \cite{Kodama:1980aa} and \cite{Harada:2015aa} for explanations) and its integral, the Kodama mass, is a conserved quantity. The Kodama mass within a sphere of radius $r$ at time $t$ is thus defined by
\begin{equation}
M_K(t,r) := 4\pi \int^r_0S^t\alpha(t,x) R^2(t,x)dx.
\end{equation}
By developing $S^t$, we find
\begin{equation}
M_K(t,r) = 4\pi\int^r_0\left[T^t_t\left(-R^2\partial_rR\right)+T^t_r\left(R^2\partial_tR\right)\right]dx.
\end{equation}
The expression \eqref{Tmunu} gives, in the case of a universe filled with one fluid of matter (other cases do not change much things),
\begin{eqnarray}
T^t_t &=& -(e+p)W^2+p = -E\\
T^t_r &=& (e+p)\frac{W^2}{\alpha}v_r = \frac{S_r}{\alpha}.
\end{eqnarray}
In terms of BSSN variables, we have
\begin{eqnarray}
R^2\partial_rR &=& \psi^6a^3r^2 \sqrt{\frac{\hat{b}}{\hat{a}}}\left(1+2r\frac{\partial_r\psi}{\psi}+\frac{r\partial_r\hat{b}}{2\hat{b}}\right)\\
R^2\partial_tR &=& \psi^6a^3r^3\sqrt{\frac{\hat{b}}{\hat{a}}}\left(\frac{\dot{a}}{a}+2\frac{\partial_t\psi}{\psi} + \frac{\partial_t\hat{b}}{2\hat{b}}\right) \nonumber\\
&=& -\alpha\psi^6a^3r^3\sqrt{\frac{\hat{b}}{\hat{a}}}\left(\frac{K}{3}+A_b\right),
\end{eqnarray}
where we have use \eqref{BSSN2} and \eqref{BSSN3} for the last equality. In conclusion, the expression for the Kodama mass in BSSN variables is
\begin{eqnarray}
M_K(t,r) &=& 4\pi a^3 \int^r_0 \psi^6 x^2\sqrt{\frac{\hat{b}}{\hat{a}}}\Bigg[ E\left(1+2r\frac{\partial_r\psi}{\psi}+\frac{r\partial_r\hat{b}}{2\hat{b}}\right)\nonumber\\
 &&- xS_r\left(\frac{K}{3}+A_b\right)\Bigg]dx.
\label{Kodama_in_BSSN_var}
\end{eqnarray}
The corresponding quantity for the Friedmann universe used as background is thus
\begin{equation}
\overline{M_K}(t,r) = 4\pi a^3\int^r_0 \overline{e} x^2dx = \frac{4}{3}\pi (ar)^3\overline{e}.
\end{equation}
Note that in the definition of the compaction function \eqref{compact_definition} we need to compute it at the same areal radius than the local Kodama mass, that is to say
\begin{eqnarray*}
&&\overline{M_K}(t,\psi^2\sqrt{\hat{b}}r) = \frac{4}{3}\pi a^3\psi^6\overline{e} \sqrt{\frac{\hat{b}}{\hat{a}}}r^3\\
&&= 4\pi a^3\overline{e}\int^{\psi^2\sqrt{\hat{b}}r}_0 x^2dx\\
&&= 4\pi a^3\overline{e} \int^r_0  \psi^6y^2 \sqrt{\frac{\hat{b}}{\hat{a}}}\left(1+2y\frac{\partial_r\psi}{\psi}+\frac{y\partial_r\hat{b}}{2\hat{b}}\right)dy,
\end{eqnarray*}
where we made the change of variable $x = \psi(t,y)^2\sqrt{\hat{b}(t,y)}y$ for the last equality. The last expression, though less simple, can be useful because of its similarity with the first term of \eqref{Kodama_in_BSSN_var}. For example, in a comoving gauge it gives the relation 
\begin{eqnarray}
M_K(t,r) - \overline{M_K}(t,\psi^2\sqrt{\hat{b}}r) = \nonumber\\
4\pi a^3 \overline{e}\int^r_0 \delta\psi^6 \sqrt{\frac{\hat{b}}{\hat{a}}}x^2\Bigg(1+2x\frac{\partial_r\psi}{\psi}+\frac{x\partial_r\hat{b}}{2\hat{b}}\Bigg)dx,
\label{comoving_dM}
\end{eqnarray}
only in term of the energy-density contrast $\delta = \frac{e-\overline{e}}{\overline{e}}$.

Concerning the mean energy-density contrast, recall that it is defined in the following way :
\begin{equation}
\delta_\text{mean}(t,r) = \frac{\int^R_0\delta R^2dR}{\int^R_0R^2dR}.
\end{equation}
By using BSSN equations, it gives
\begin{equation}
\delta_\text{mean}(t,r) = \frac{\int^r_0 \delta \psi^6a^3 \sqrt{\frac{\hat{b}}{\hat{a}}}x^2\left(1+2x\frac{\partial_r\psi}{\psi}+\frac{x\partial_r\hat{b}}{2\hat{b}}\right)dx}{\int^r_0\psi^6a^3 \sqrt{\frac{\hat{b}}{\hat{a}}}x^2\left(1+2x\frac{\partial_r\psi}{\psi}+\frac{x\partial_r\hat{b}}{2\hat{b}}\right)dx},
\end{equation}
where the denominator can be replaced by $\frac{R^3}{3} = \frac{\psi^6a^3\sqrt{\frac{\hat{b}}{\hat{a}}}r^3}{3}$. We thus see the direct relation between the mean energy-density contrast and the compaction in the comoving gauge by looking at the relation \eqref{comoving_dM}.

\section*{Appendix D : Derivation of the long-wavelength solution in the BSSN variables}

We assume that, at fixed time, the universe becomes locally flat homogeneous and isotropic in the limit $\epsilon \to 0$. We thus assume, still following \cite{Lyth:2005aa}, that $\hat{\gamma}_{ij} = O(\epsilon^2)$. As described in details in \cite{Harada:2015aa}, we then obtain
\begin{eqnarray}
\psi &=& O(\epsilon^0),\\
v^i &=& O(\epsilon),\\
v_i &=& O(\epsilon^3),\label{v_i_long_wavelength}\\
D_i v^i &=& O(\epsilon^4),\\
W &=& 1+O(\epsilon^6),\\
\delta &=& O(\epsilon^2),\\
\hat{A}_{ij} &=& O(\epsilon^2),\\
h_{ij} &=& O(\epsilon^2),\\
\chi &=& O(\epsilon^2),\\
\kappa &=& O(\epsilon^2),
\end{eqnarray}
where we used the following notations
\begin{eqnarray*}
\delta &:=& \frac{e-\overline{e}}{\overline{e}},\\
h_{ij} &:=& \hat{\gamma}_{ij} - \overline{\hat{\gamma}}_{ij},\\
\chi &:=& \alpha - 1,\\
\kappa &:=& \frac{K-\overline{K}}{\overline{K}}.
\end{eqnarray*}

Moreover, the conformal factor can be decomposed in such a way :
\begin{equation}
\psi(t,x^i) = \Psi(x^i)\left(1+\xi(t,x^i)\right),
\end{equation}
where $\Psi = O(\epsilon^0)$ and $\xi = O(\epsilon^2)$.  In fact, as also shown in \cite{Lyth:2005aa} and \cite{Harada:2015aa}, all slicings coincide up to $O(\epsilon)$, which allows to apply the long-wavelength scheme in any slicing. After having done it in the special case where $p = \omega e$, the evolution equations become
\begin{eqnarray}
&&\dot{\delta} + 6 \dot{\xi} + 3H\omega\left(\chi + \kappa\right) = O(\epsilon^4),\label{eq_approx1}\\
&&\frac{1}{1+\omega}\dot{\delta}+6\dot{\xi} = O(\epsilon^4),\label{eq_approx2}\\
&&\partial_t\left(a^3(1+\omega)\overline{e}u_i\right) = -a^3\overline{e}\left[ \omega\partial_i \delta + (1+\omega )\partial_i \chi\right] \nonumber\\
&&\text{\ \ \ \ \ \ \ \ \ \ \ \ \ \ \ \ \ \ \ \ \ \ }+ O(\epsilon^5),\label{eq_approx3}\\
&&\overline{\Delta}\Psi = -2\pi \Psi^5a^2\overline{e}(\delta-2\kappa) + O(\epsilon^4),\label{eq_approx4}\\
&&H^{-1}\dot{\kappa} = \frac{3\omega-1}{2}\kappa - \frac{3(1+\omega)}{2}\chi -\frac{1+3\omega}{2}\delta\nonumber\\
 &&\text{\ \ \ \ \ \ \ \ \ \ \ \ \ \ \ \ \ \ \ \ \ \ }+ O(\epsilon^4),\label{eq_approx5}\\
&&\partial_t h_{ij} = -2 \hat{A}_{ij} + O(\epsilon^4),\label{eq_approx6}\\
&&\partial_t\hat{A}_{ij} + 3H\hat{A}_{ij} = \frac{1}{a^2\Psi^4}\Bigg[-\frac{2}{\Psi}\left(\overline{D}_i\overline{D}_j\Psi - \frac{1}{3}\overline{\hat{\gamma}}_{ij}\overline{\Delta}\Psi\right)\nonumber \\
 &&+ \frac{6}{\Psi^2}\left(\overline{D}_i\Psi\overline{D}_j\Psi - \frac{1}{3}\overline{\hat{\gamma}}_{ij}\overline{D}^k\Psi\overline{D}_k\Psi\right)\Bigg]+ O(\epsilon^4),\label{eq_approx7}\\
&&\overline{D}_i\left(\Psi^6\hat{A}^i_j\right) +2H\Psi^6\overline{D}_j\kappa = 8\pi\Psi^6(1+\omega)\overline{e}u_j\nonumber\\
 &&\text{\ \ \ \ \ \ \ \ \ \ \ \ \ \ \ \ \ \ \ \ \ \ } +O(\epsilon^5),\label{eq_approx8}
\end{eqnarray}
where $\overline{D}$ and $\overline{\Delta}$ are the covariant derivative and the Laplacian operators related to the flat metric written in the coordinates $\lbrace x^i\rbrace$. In spherical coordinates, we have that this metric is given by $\overline{\hat{\gamma}}_{ij} = \text{diag}(1,r^2,r^2\sin^2\theta)$. The resulting equations have been solved in \cite{Harada:2015aa} for the constant mean curvature (CMC) slicing (for which $K = \overline{K}$), the comoving slicing, the uniform-density slicing (for which $\delta = 0$) and the geodesic slicing. The paper also gives comparison of the solutions between these four slicings. In our case, we will solve them for a general Bona-Masso slicing \eqref{slicing}. This solution has, to our knowledge, never been derived in the literature.

We start with the computation of the variables $\hat{A}_{ij}$ and $h_{ij}$ which, as explained in \cite{Harada:2015aa}, do not depend on the slicing to $O(\epsilon^2)$. If we use the intermediate variable
\begin{eqnarray}
p_{ij} &:=& \frac{1}{\Psi^4}\Bigg[-\frac{2}{\Psi}\left(\overline{D}_i\overline{D}_j\Psi - \frac{1}{3}\overline{\hat{\gamma}}_{ij}\overline{\Delta}\Psi\right)\nonumber\\
 &&+\frac{6}{\Psi^2}\left(\overline{D}_i\Psi\overline{D}_j\Psi - \frac{1}{3}\overline{\hat{\gamma}}_{ij}\overline{D}^k\Psi\overline{D}_k\Psi\right)\Bigg],
\end{eqnarray}
the explicit expressions of $\hat{A}_{ij}$ and $h_{ij}$ are given, solving \eqref{eq_approx6} and \eqref{eq_approx7}, by
\begin{eqnarray}
&&\hspace*{-10mm}\hat{A}_{ij} = \frac{2}{3\omega +5}p_{ij} H \left(\frac{1}{aH}\right)^2 + O(\epsilon^4),\\
&&\hspace*{-10mm}h_{ij} = -\frac{4}{(3\omega +5)(3\omega+1)}p_{ij}\left(\frac{1}{aH}\right)^2+ O(\epsilon^4).
\end{eqnarray}
If we want to use spherical symmetry and the BSSN variables presented in section \ref{evolution eq}, we must use the quantities
\begin{eqnarray}
p_a &:=& p^r_r = \frac{1}{\Psi^4}\Bigg[-\frac{4}{3\Psi}\left(\partial^2_r\Psi -\frac{1}{r}\partial_r\Psi\right)\nonumber\\
&& \hspace*{2cm}+ \frac{4}{\Psi^2}\left(
\partial_r\Psi\right)^2\Bigg],\\
p_b &:=& p^\theta_\theta =  \frac{1}{\Psi^4}\Bigg[\frac{2}{3\Psi}\left(\partial^2_r\Psi -\frac{1}{r}\partial_r\Psi\right)\nonumber\\
&& \hspace*{2cm} - \frac{2}{\Psi^2}\left(
\partial_r\Psi\right)^2\Bigg].
\end{eqnarray}
With these definitions, we can write 
\begin{eqnarray*}
A_a &=& \frac{2}{3\omega +5}p_a H \left(\frac{1}{aH}\right)^2 + O(\epsilon^4),\\
A_b &=& \frac{2}{3\omega +5}p_b H \left(\frac{1}{aH}\right)^2 + O(\epsilon^4),\\
\hat{a} &=& 1 - \frac{4}{(3\omega +5)(3\omega+1)}p_a\left(\frac{1}{aH}\right)^2 + O(\epsilon^4),\\
\hat{b} &=& 1 - \frac{4}{(3\omega +5)(3\omega+1)}p_b\left(\frac{1}{aH}\right)^2 + O(\epsilon^4),
\end{eqnarray*}
where we have $p_a+2p_b = 0$, implying the desired conditions $A_a + 2A_b = 0$ and $\hat{a}\hat{b}^2 = 1 + O(\epsilon^4)$. The variable $\hat{\Delta}^r$ can be determined thanks to equation \eqref{Delta}.

We will now determine the expressions for $\delta$, $\xi$, $\kappa$ and $\chi$ by solving equations \eqref{eq_approx1}, \eqref{eq_approx2}, \eqref{eq_approx4}, \eqref{eq_approx5} and \eqref{slicing}. The combination of the first two ones gives
\begin{equation}
H^{-1}\dot{\delta} + 3(1+\omega)(\chi + \kappa) = O(\epsilon^4).
\end{equation}
The Bona-Masso equation \eqref{slicing} reads, after being expanded around $\alpha = 1$,
\begin{equation}
H^{-1}\dot{\chi} = 3f(1) \kappa +O(\epsilon^4).
\end{equation}
With equation \eqref{eq_approx5} and the change of variable $s = \ln a$, we obtain the differential linear system 
\begin{eqnarray}
\partial_s \begin{pmatrix}
\kappa\\
\delta\\
\chi
\end{pmatrix} = M(\omega) \begin{pmatrix}
\kappa\\
\delta\\
\chi
\end{pmatrix} + O(\epsilon^4),\label{differential_system}
\end{eqnarray}
where the matrix $M(\omega)$ is given by
\begin{eqnarray}
\displaystyle M(\omega) = \displaystyle\begin{pmatrix}
\displaystyle\frac{3\omega-1}{2} & \displaystyle -\frac{1+3\omega}{2} & \displaystyle -\frac{3(1+\omega)}{2}\\
\displaystyle -3(1+\omega) & \displaystyle 0 & \displaystyle -3(1+\omega)\\
\displaystyle 3f(1) & \displaystyle 0 & \displaystyle 0
\end{pmatrix}.
\end{eqnarray}
This matrix admits the following eigenvalues :
\begin{eqnarray}
&&\hspace*{-5mm}\lambda_* = 1+3\omega,\\
&&\hspace*{-5mm}\lambda_\pm = -\frac{3}{4}\left[ 1+\omega \pm \sqrt{(1+\omega)(1+\omega-8f(1))}\right].
\end{eqnarray}

The general solution of the system \eqref{differential_system} is thus given by
\begin{eqnarray}
\hspace*{-4mm}\begin{pmatrix}
\kappa\\
\delta\\
\chi
\end{pmatrix} \hspace*{-1mm}= C_* \mathbf{v}_* a^{\lambda_*} + C_+ \mathbf{v}_+ a^{\lambda_+} + C_- \mathbf{v}_- a^{\lambda_-}  + O(\epsilon^4),\label{general_solution}
\end{eqnarray}
where $\mathbf{v}_*$, $\mathbf{v}_+$ and $\mathbf{v}_-$ are the eigenvectors associated respectively to $\lambda_*$, $\lambda_+$ and $\lambda_-$ and $C_*$, $C_+$ and $C_-$ are integration constants.

For the geodesic slicing, $\lambda_+ = \displaystyle -\frac{3(1+\omega)}{4} <0$ and $\lambda_- = 0$. In the other Bona-Masso slicings of Table \ref{slicing_conditions}, $\lambda_+$ and $\lambda_-$ are complex conjugates with a negative real part. Since we only take pure growing modes, we set $C_+ = C_-  = 0$. We have that the eigenvector of $\lambda_*$ is given by
\begin{eqnarray}
\begin{pmatrix}
v_*^1\\
v_*^2\\
v_*^3
\end{pmatrix} = \begin{pmatrix}
(1+3\omega )^2\\
-3(1+\omega)(1+3\omega+3f(1))\\
3(1+3\omega)f(1)
\end{pmatrix}.
\end{eqnarray}

We can then deduce the two useful relations
\begin{eqnarray}
\kappa &=& \frac{v_*^1}{v_*^2}\delta + O(\epsilon^4),\label{kappa_v_delta}\\
\chi &=&  \frac{v_*^3}{v_*^2}\delta + O(\epsilon^4).\label{chi_v_delta}
\end{eqnarray}

By inserting \eqref{kappa_v_delta} in \eqref{eq_approx4}, we finally obtain 
\begin{equation}
\delta = \frac{v_*^2}{v_*^2-2v_*^1}F \left(\frac{1}{aH}\right)^2 + O(\epsilon^4),
\end{equation}
where we have defined 
\begin{equation}
F := -\frac{4}{3} \frac{\overline{\Delta \Psi}}{\Psi^5}.
\end{equation}
We immediately deduce the values of $\kappa$, $\chi$ and $\xi$ thanks to \eqref{kappa_v_delta}, \eqref{chi_v_delta} and \eqref{eq_approx2} :
\begin{eqnarray}
\kappa &=&  \frac{v_*^1}{v_*^2-2v_*^1}F \left(\frac{1}{aH}\right)^2 + O(\epsilon^4),\\
\chi &=&  \frac{v_*^3}{v_*^2-2v_*^1}F \left(\frac{1}{aH}\right)^2 + O(\epsilon^4),\\
\xi &=& -\frac{1}{6(1+\omega )} \frac{v_*^2}{v_*^2-2v_*^1}F \left(\frac{1}{aH}\right)^2 + C\nonumber\\
&&\hspace*{3cm} + O(\epsilon^4),
\end{eqnarray}
where the integration constant $C$ can be absorbed into $\Psi$ (see \cite{Harada:2015aa} in a similar case).

The last equation that remains to solve is \eqref{eq_approx3} to find the velocity $v_i$. By using the expressions of $\delta$ and $\chi$, we find
\begin{eqnarray}
u_i &=& -\frac{2}{3\omega+5}\cdot\frac{\omega v_*^2+(1+\omega) v_*^3}{(1+\omega)(v_*^2-2v_*^1)}\partial_i F \nonumber\\
&&\cdot\frac{1}{a}\left[\int^a_0 \frac{d\tilde{a}}{H(\tilde{a})} +C\right]\left(\frac{1}{aH}\right)^2 + O(\epsilon^5),
\end{eqnarray}
where the integration constant $C$ must be set to zero to keep only growing modes (see \cite{Harada:2015aa}). We deduce the value for $v_i$ : 
\begin{equation}
v_i = -\frac{2}{3\omega+5}\cdot\frac{\omega v_*^2+(1+\omega) v_*^3}{(1+\omega)(v_*^2-2v_*^1)}\partial_i F a \left(\frac{1}{aH}\right)^3 + O(\epsilon^5).
\end{equation}

To summarize the results in terms of BSSN variables in spherical symmetry, the long-wavelength solution is given by
\begin{eqnarray*}
\psi &=& \Psi \left(1 -\frac{1}{6(1+\omega )} \frac{v_*^2}{v_*^2-2v_*^1}F \left(\frac{1}{aH}\right)^2\right)\nonumber\\
&&\hspace*{3cm} + O(\epsilon^4),\\
A_a &=& \frac{2}{3\omega +5}p_a H \left(\frac{1}{aH}\right)^2 + O(\epsilon^4),\\
A_b &=& \frac{2}{3\omega +5}p_b H \left(\frac{1}{aH}\right)^2 + O(\epsilon^4),\\
\hat{a} &=& 1 - \frac{4}{(3\omega +5)(3\omega+1)}p_a\left(\frac{1}{aH}\right)^2 + O(\epsilon^4),\\
\hat{b} &=& 1 - \frac{4}{(3\omega +5)(3\omega+1)}p_b\left(\frac{1}{aH}\right)^2 + O(\epsilon^4),\\
\delta &=& \frac{v_*^2}{v_*^2-2v_*^1}F \left(\frac{1}{aH}\right)^2 + O(\epsilon^4),\\
K &=& \overline{K}\left(1+ \frac{v_*^1}{v_*^2-2v_*^1}F \left(\frac{1}{aH}\right)^2\right) +O(\epsilon^4),\\
\alpha &=& 1+\frac{v_*^3}{v_*^2-2v_*^1}F \left(\frac{1}{aH}\right)^2 + O(\epsilon^4),\\
v_r &=& -\frac{2}{3\omega+5}\cdot\frac{\omega v_*^2+(1+\omega) v_*^3}{(1+\omega)(v_*^2-2v_*^1)}\partial_i F a \left(\frac{1}{aH}\right)^3\nonumber\\
&&\hspace*{3cm} + O(\epsilon^5).
\end{eqnarray*}

\bibliography{./biblio}

\end{document}